\documentclass[prx,twocolumn,showpacs,amsmath,amssymb]{revtex4-1}
\usepackage{graphicx,amsmath,bm, braket, color} 
\usepackage[final]{pdfpages}
\pdfoutput=1

\begin{document}
\title {Cavity quantum electrodynamics with mesoscopic topological superconductors}
\author{Olesia Dmytruk, Mircea Trif and Pascal Simon}
\address{Laboratoire de Physique des Solides, CNRS UMR-8502, Universit\'{e} Paris Sud, 91405 Orsay cedex, France}
\date{\today}

\begin{abstract}
We study  one-dimensional $p$-wave superconductors  capacitively coupled to a microwave stripline cavity. By probing the light exiting from the cavity, one can reveal the electronic susceptibility of the  $p$-wave superconductor. We analyze two superconducting systems: the prototypical Kitaev chain, and a topological semiconducting wire. For both systems, we show that the photonic measurements, via the electronic  susceptibility,  allows us to determine the topological phase transition point, the emergence of the Majorana fermions, and the parity of their ground state. We show that all these effects, which are absent in effective theories that take into account the coupling of light to Majorana fermions only, are due to the interplay between the Majorana fermions and the bulk states of the superconductors.  
\end{abstract}


\maketitle

\section{Introduction}

 Condensed matter systems are an endless resource of emergent physical phenomena and associated quasiparticles. Majorana fermions, which are particles that are their own antiparticles and which have been first proposed as particles in the context of high energy physics, emerge beautifully as zero energy excitations in condensed matter setups  \cite{kitaev2001unpaired,alicea2012new}. Specifically, they are predicted to occur as zero energy excitations in solid-state systems, such as genuine $p$-wave superconductors  \cite{read2000paired,ivanov2001non,sato}, or  engineered from topological insulators \cite{fu2008superconducting}, semiconductor wires in a magnetic field \cite{oreg2010helical,lutchyn2010majorana, mourik2012signatures}, or in chains of magnetic atoms  \cite{nadj2013proposal,pientka2013topological,klinovaja2013topological, braunecker2013interplay, vazifeh2013self,kim2014helical,nadj2014observation}, all in the proximity of $s$-wave superconductors. These exotic objects are robust against local perturbations and, moreover, they obey non-Abelian statistics \cite{ivanov2001non,nayak2008non, alicea2011non} under braiding operations, thus recommending them as qubits for the implementation of topological quantum computation.

Electronic transport is the foremost experimental tool for investigating the Majorana fermions physics but alternative, {\it non-invasive}, methods that preserve the quantum states would be highly desired to address these objects. Cavity quantum electrodynamics (cavity QED) has been established as an extremely versatile tool to address equilibrium and out-of-equilibrium electronic and spin systems non-invasively \cite{wallraff2004strong, blais2004cavity, majer2007coupling, dicarlo2009demonstration, trif2008spin,delbecq2011coupling, frey2012dipole, petersson2012circuit, putz2014protecting}. Majorana fermions, too, have been recently under theoretical scrutiny in the context of cavity QED physics \cite{hassler2011transmon,trif2012resonantly,schmidt2013majorana,pekker2013,cottet2013squeezing,blais2013,hyart2013flux,ohm2014majorana,Yavilberg,ohm2015}. However, most of these studies dealt with effective low energy models that involved Majorana fermions only, leaving the bulk physics, which is at the heart of the Majorana physics, largely unexplored.

\begin{figure}[t] 
\centering
\includegraphics[width=0.95\linewidth]{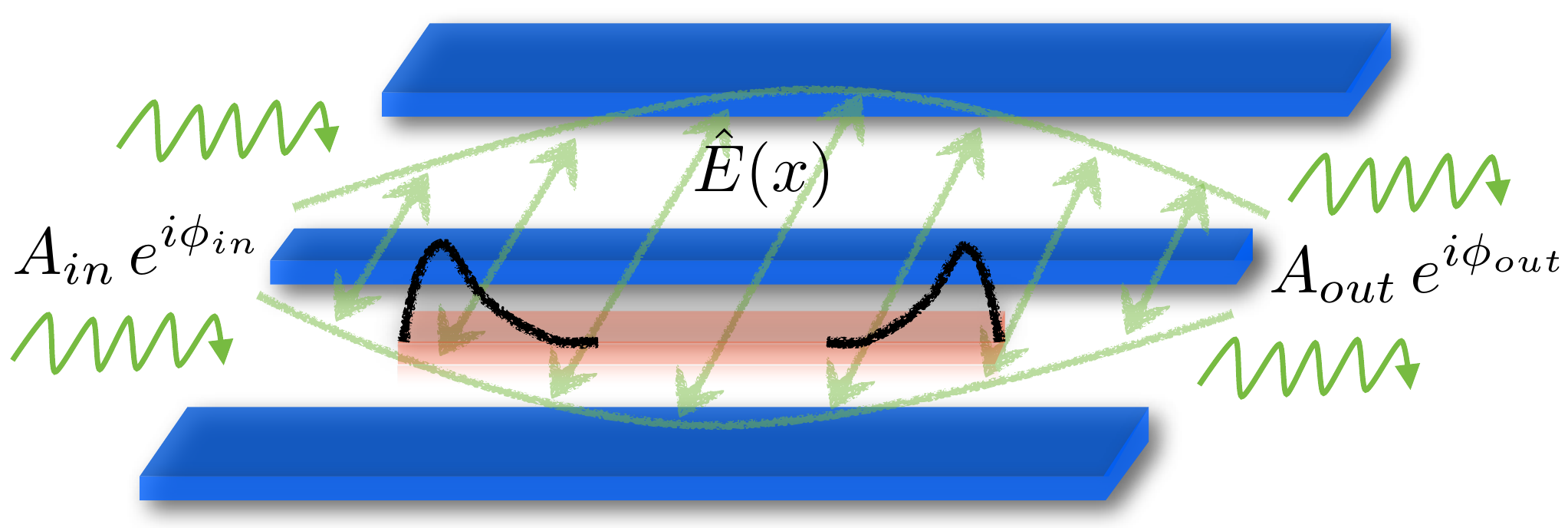}
\caption{A sketch of the system: a one dimensional system (red rectangle) is placed at the maximum of the electrical field (green straight arrows) inside a superconducting microwave cavity (blue). The electromagnetic field inside the  cavity is probed by sending input fields of amplitude and phase $A_{in}$ and $\phi_{in}$, respectively, and measuring the field at the end with $A_{out}$ and $\phi_{out}$. The difference between the two gives a direct access to the electronic correlation function in the wire (see text). The presence of Majorana end modes in the finite wire (black curves) is also signaled in the cavity response.} 
\label{fig:scheme} 
\end{figure}

The basic idea behind cavity  QED with  electronic system is that it allows one to extract various properties of the latter, such as its spectrum and its electronic distribution function, from photonic transport measurements, as opposed to electronic transport. Such photonic transport is quantified by the complex transmission coefficient  $\tau = A\exp(i\phi)$ that relates the output and input photonic fields as depicted in Fig.~\ref{fig:scheme}. In the weakly coupled limit, one finds \cite{cottet2011admittance, schiro2014tunable}, Appendix~\ref{inputoutput}: 
\begin{align}
\tau(\omega) = \frac{\displaystyle \kappa}{\displaystyle -i(\omega - \omega_c) +  \kappa - i\,\Pi(\omega)}\,,
\label{transmissioncoef}
\end{align}
where $\omega_c$ and $\kappa$ are the frequency and the escape rate of the cavity, respectively, while $\Pi(\omega)$ is an electronic correlation function that depends on the actual coupling between the two systems, and which contains  information about the spectrum of the electronic system. The phase and amplitude response of the cavity close to resonance $\omega\approx\omega_c$ are related to the susceptibility $\Pi(\omega)$ as follows: $\delta\phi=\Pi'(\omega)/\kappa$ and $\delta A/A_{in}=\Pi''(\omega)/\kappa$, where $\delta\phi=\phi_{out}-\phi_{in}$, $\delta A=A_{in}-A_{out}$,  and $\Pi'(\omega)={\mathcal Re}[\Pi(\omega)]$ ($\Pi''(\omega)={\mathcal Im}[\Pi(\omega)]$) is the real (imaginary) part of the susceptibility. 

In this paper, we evaluate the function $\Pi(\omega)$ first for the  simple case of a one-dimensional (1D) $p$-wave superconductor described by the Kitaev chain and then for more realistic model of a 1D topological semiconducting wire in proximity of a superconductor. We assume in both cases that these 1D systems are  coupled to a microwave cavity, as showed schematically in  Fig.~\ref{fig:scheme}.   We address  various physical  situations for this coupling and show that such a method allows us to ascertain the topological phase transition point, the occurrence of Majorana fermions, and the parity of the ground state, all in a {\it global} and {\it non-invasive} fashion.  
The paper is structured as follows. In Sec.~\ref{sec2}, we describe our model Hamiltonians for the two systems under 
consideration and discuss the coupling between the microwave photons and the electrons in the 1D topological systems.
In Sec.~\ref{sec3}, we show how the optical transmission through the cavity is able to probe the topological phase transition. In~\ref{sec4}, we demonstrate that the cavity allows to detect the occurrence of Majorana fermions and the parity of the Majorana
fermionic state in a non-invasive fashion. Finally, in Sec.~\ref{sec5} we provide a brief summary of our results. Technical details of the calculations are given in the appendices.

\section{Model Hamiltonian}\label{sec2}

In the following we will consider various models of $p$-wave superconductors coupled to a microwave (superconducting) cavity, such as the Kitaev $p$-wave superconductor model, and the spin-orbit coupled nanowire subjected to a magnetic field and in the proximity of an $s$-wave superconductor.    

The general Hamiltonian for the one-dimensional systems we consider here is of the form:
\begin{equation}
H_{sys}=H_{el}+H_{el-c}+H_{ph}\,,
\label{totalhamiltonian}
\end{equation}
being the sum of the electronic Hamiltonian, its capacitive coupling to the cavity, and the free photon field, respectively. While the  electronic term is model specific, and it will be discussed below, the last two terms read:
\begin{equation}
H_{el-c}= \alpha\sum\limits_{i = 1}^{N}\hat{n}_i\,(a + a^\dag)\,,
\label{interactionhamiltonian}
\end{equation}
and 
\begin{equation}
H_{ph}= \omega_c a^\dag a\,.\label{cavityhamiltonian}
\end{equation}
In Eq. \eqref{interactionhamiltonian},  $a^\dag (a)$ is the photon creation (annihilation) operator, respectively. $\alpha$ is the electron-photon coupling constant that couples to the charge density $\hat{n}$. This merely acts as to  shift the chemical potential.
 In Eq. \eqref{cavityhamiltonian}, $\omega_c$ is the frequency of the  photonic mode (setting $\hbar = 1$ throughout). Such a model could be realized experimentally by coupling a spin-orbit nanowire in the presence of a Zeeman field to a nearby $s$-wave superconductor \cite{lutchyn2010majorana, oreg2010helical}.  In the present setup, which is based on a microwave superconducting stripline cavity,  the $s$-wave superconductor that induces superconducting correlations in the wire could be a part of the underlaying cavity.  For example, the nanowire could be tunnel-coupled to the central superconducting material showed in Fig.~\ref{fig:scheme}.  We have a considered a global capacitive coupling between the electronic nanosystem and the cavity electric field. Such a coupling can be justified by a full microscopic approach (see Appendix \ref{sec:effective} for details and also Ref. \cite{cottet2015electron} that provides a microscopic  description of the electric coupling between electrons in a nanocircuit and cavity photons). 

By solving the equation of motion $d a/dt=-i[a,H_{sys}]$ for the photonic field iteratively up to second order in $\alpha$ with respect to the cavity frequency $\omega_c$~\cite{schiro2014tunable}, we find for the correlation function $\Pi(\omega)$ in Eq.~\eqref{transmissioncoef} in the time domain
\begin{equation}
\Pi(t-t')=-i\alpha^2\theta(t-t')\langle [\hat{n}_{I}(t),\hat{n}_I(t')]\rangle\,,
\label{susc_time}
\end{equation}
being the {\it total} charge susceptibility of the electronic system (which can be here a  1D $p-$wave superconductor or a topological 1D wire). In Eq.~\eqref{susc_time}, we introduced $\hat{n}_I(t)=U^\dagger(t)\hat{n}U(t)$, with $\hat{n}$ being the total number of electrons operator and $U(t)=\exp{(-iH_{el}t)}$ the evolution operator for the electronic system. We assume zero temperature limit ($T=0$) so that the average $\langle\dots\rangle$ is taken over the superconducting ground state. Note that $\Pi(\omega)=\int_{-\infty}^\infty dt\exp{(i\omega t)}\Pi(t)$ and that $\Pi(\omega)\equiv 0$ in the absence of superconductivity ($\Delta=0$), i.e. there are no effects from such a coupling for a wire in the normal state.   We detail below the models for both
topological 1D systems we consider in this paper.

\subsection{Kitaev chain}

The simplest model of a $p$-wave superconductor that hosts Majorana fermions is  the  Kitaev chain \cite{kitaev2001unpaired}. Therefore, we first consider for the electronic part in Eq. \eqref{totalhamiltonian}, the Kitaev Hamiltonian $H_{el}^K$ that  reads:
\begin{align}
\!\!\!\!\!H_{el}^K =\!- \mu\sum\limits_{j = 1}^{N}c^\dag_jc_j& -\!\frac{1}{2}\sum\limits_{j = 1}^{N - 1}({\rm t}\,c^\dag_jc_{j + 1} \!+\Delta\,c_jc_{j + 1} \!+ {\rm h.c.})\,,
\label{kitaevhamiltonian}
\end{align}
where ${\rm t}$ is the hopping parameter, $\Delta$ is the $p$-wave superconducting pairing potential, $\mu$ is the chemical potential, and $N$ is the total number of sites. Also,  $c_j^\dag (c_j)$ is the creation (annihilation) electronic operator at the site $j$. Note that the electronic operators are spinless, and the electronic density is given by $\hat{n}=\sum_{j=1}^Nc_j^\dagger c_j$.  
In the present setup, which is based on a microwave superconducting stripline cavity,  the $s$-wave superconductor that induces superconducting correlations in the wire could be a part of the underlaying cavity.  For example, the nanowire could be tunnel-coupled to the central (super-)conductor showed in Fig.~\ref{fig:scheme}. The fact that microwave photons effectively couple only to electrons of the Kitaev chain is accounted for  in Appendix \ref{sec:effective}.

The Kitaev Hamiltonian in Eq.~\eqref{kitaevhamiltonian} can be easily  diagonalized. The susceptibility in this case can be found by simply substituting the expression for  the density in Eq.~\eqref{susc_time} with the one corresponding to the Kitaev  model.  We will discuss its physical content in Sec.~\ref{sec3} and Sec.~\ref{sec4}.

\subsection{Spin-orbit coupled nanowire}

A realistic system that can emulate, in some limits, the Kitaev chain consists of a nanowire with a spin-orbit interaction, subjected to an external  magnetic field, and coupled by proximity effect to an $s$-wave superconductor \cite{oreg2010helical,lutchyn2010majorana, mourik2012signatures}. The entire system is then assumed to be (capacitively) coupled to the microwave cavity. The tight-binding Hamiltonian $H_{el}^W$  for the nanowire with spin-orbit (SO) interaction  in the presence of the magnetic field reads \cite{rainis2013towards} 
\begin{align}
H_{el}^W&=-{\rm t}\sum_{j=1}^{N-1}c^\dag_{j+1\alpha}\delta_{\alpha\beta}c_{j\beta}-\mu\sum_{j=1}^{N}c^\dag_{j\alpha}\delta_{\alpha\beta}c_{j\beta}\nonumber\\
&+\Delta\sum_{j=1}^{N}c^\dag_{j\uparrow}c^\dag_{j\downarrow}-i\gamma\sum_{j=1}^{N-1}c^\dag_{j+1\alpha}\sigma^y_{\alpha\beta}c_{j\beta}\nonumber\\
&-V_Z\sum_{j=1}^{N}c^\dag_{j\alpha}\sigma^x_{\alpha\beta}c_{j\beta}+{\rm h. c.}\,,
\end{align}
where, as before, ${\rm t}$ and $\mu$ are the hopping amplitude and the chemical potential, respectively, $\gamma$ is the spin-flip hopping amplitude (or the spin-orbit coupling), $\Delta$ is the $s$-wave pairing  potential induced by proximity, $V_Z$ is the Zeeman splitting energy ($V_Z=-g\mu_BB/2$, with $g$ and $B$ being the $g$-factor and external magnetic fields, respectively). Also, $c_{j\sigma}$ ($c^\dagger_{j\sigma}$) are the annihilation (creation) operators for electrons at site $j$ and spin $\sigma=\uparrow,\downarrow$, and $\sigma_i$, with $i=x,y,z$ are the Pauli matrices that act in the spin space. This model accounts thus  for spinfull electrons. Note that we assumed the spin-orbit field and the magnetic field to be orthogonal. The coupling to the cavity is again capacitive, and the density reads in this case $\hat{n}_{i}=\sum_{\sigma}c_{i\sigma}^\dagger c_{i\sigma}$. In order to find the susceptibility, we need to substitute this expression for the electronic density in Eq.~\eqref{susc_time}, and we will discuss the various cases in the following sections.   

\section{Topological phase transition}\label{sec3}

Next we will show that the topological phase transition can be inferred from the cavity response from the transmission $\bm{\tau}(\omega)$ via the susceptibility  $\Pi(\omega)$.  This function  can be calculated straightforwardly  in the case of a closed ring, i.e. for periodic  boundary conditions (PBCs), so that $c_{N + 1}\equiv c_1$ for the Kitaev chain ($c_{N + 1\sigma}\equiv c_{1\sigma}$ for the SO nanowire).  

\subsection{Kitaev chain}

\begin{figure}[t] 
\includegraphics[width=0.95\linewidth]{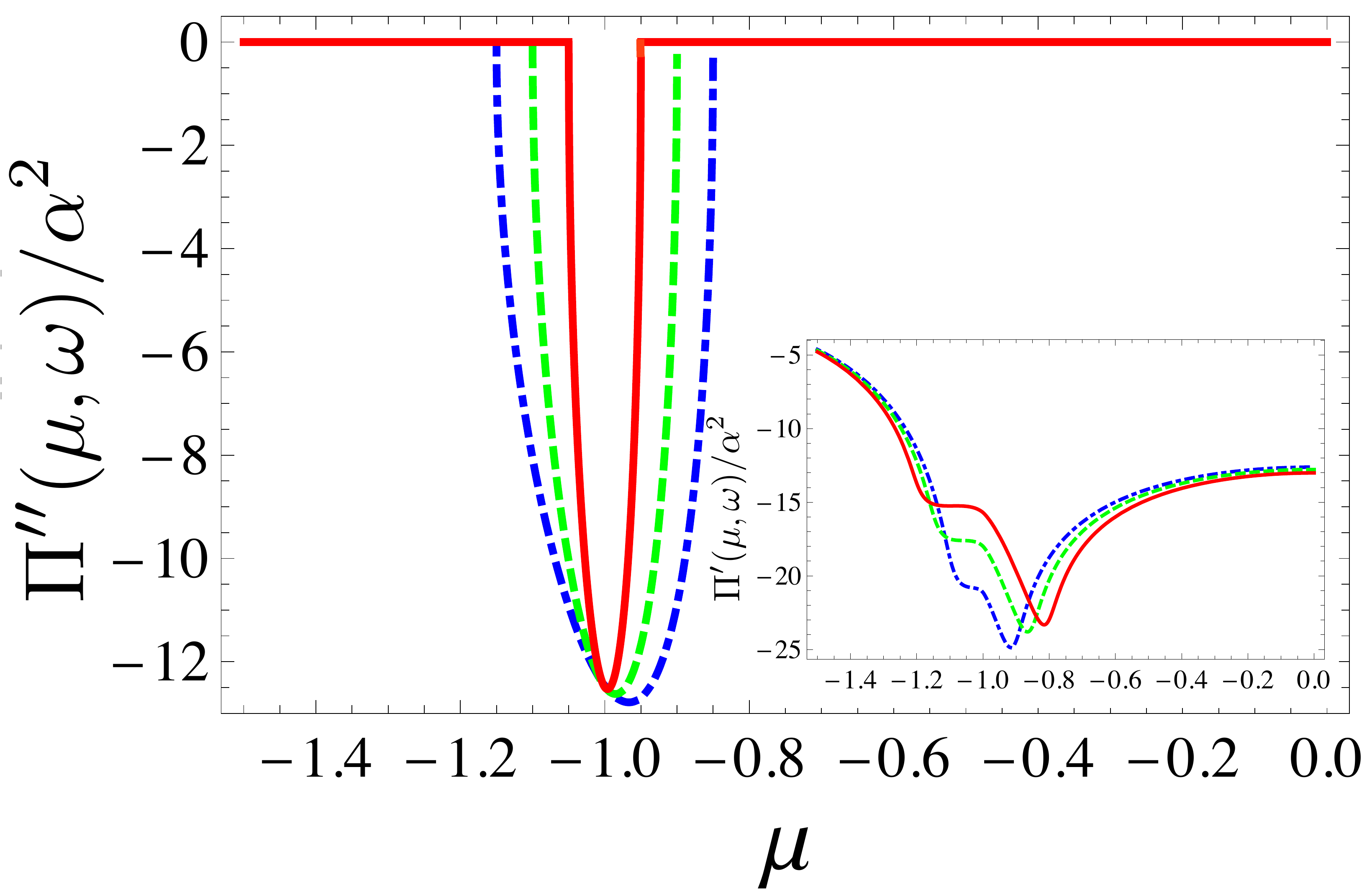}
\caption{The imaginary part of the density-density correlation function [$\Pi''(\omega)$] as a function of  $\mu$ for the Kitaev model.  The topological phase transition takes place at $\mu=-1$, where this function reaches its maximum, indicating the transition point. Inset: The real part of the density-density correlation function [$\Pi'(\omega)$], which also shows features (kinks) around the topological phase transition point. The full, dashed, dot-dashed curves correspond to the $\omega=0.2$, $0.3$, and $0.4$, respectively. We assumed  ${\rm t} = \Delta = 1$, $N = 50$ and all energies are expressed in terms of ${\rm t}$.} 
\label{topo_trans} 
\end{figure}

For PBCs, we can utilize the Fourier description for the electronic operators: $c_{j}=1/\sqrt{N}\sum_{k}e^{ikj}c_k$, with  $k=2\pi n/N$ (assuming the lattice spacing $d\equiv1$ thereon), with $n=1\dots N$. For more details see Appendix~\ref{bulksusceptibility}.  By doing so, we can readily write down the electronic Hamiltonian $H_{el}^K=\sum_{k}H_{BdG}^K(k)$, with 
\begin{equation}
H_{BdG}^K(k)=(-{\rm t}\cos{k}-\mu)\,\tau_z^k-\Delta\sin{k}\,\tau_y^k\,,
\label{BdG}
\end{equation}
where $\vec{\tau}^{k}=(\tau_x^k,\tau_y^k,\tau_z^k)$ are Pauli matrices that act in the Nambu (particle-hole) space, i.e. on the vectors $\vec{c}_k=(c_k,c_{-k}^\dagger)$. The coupling to the cavity, on the other hand, simply reads
\begin{equation}
H_{el-c}=\alpha\sum_{k}\tau_z^k(a^\dagger+a)\,,
\label{ElCav}
\end{equation} 
so that the susceptibility in the time domain can be written as:
\begin{equation}
\Pi(t)=-i\alpha^2\sum_{k}\langle0|[\tau_z^k(t),\tau_z^k(0)]|0\rangle\,,
\end{equation}
with $\tau_z^k(t)=e^{iH^K_{BdG}(k)t}\tau_z^ke^{-iH^K_{BdG}(k)t}$. Utilizing this description, after some lengthy but straightforward calculations, we obtain for the susceptibility (in the $\omega$ space): 
\begin{align}
\Pi(\omega)&=-\alpha^2\sum_{k > 0;p=\pm} \frac{\displaystyle \left(\Delta\sin k\right)^2}{\displaystyle E_{k}^2}\frac{\displaystyle p}{\displaystyle \omega + 2pE_{k} + i\eta}\,,
\end{align}
where $E_k=\sqrt{(-{\rm t}\cos{k}-\mu)^2+(\Delta\sin{k})^2}$ is the Bogoliubov spectrum of the 1D $p$-wave superconductor [from diagonalizing the BdG Hamiltonian in Eq.~\eqref{BdG}] \cite{kitaev2001unpaired} and $\eta$ is a small positive number that accounts for causality. 
For ${\rm t}=\Delta$, the imaginary part $\Pi''(\omega)$ acquires a simple analytical form, and it is given by
\begin{align}
\Pi''(\omega) = \frac{\alpha^2 {\rm t} N}{2\mu\omega}\sqrt{1 - \frac{\left[\left(\omega/2\right)^2 - {\rm t}^2 - \mu^2\right]^2}{\displaystyle 4{\rm t}^2\mu^2}}\,,
\end{align}
for $\left|{\rm t} + \mu\right| < \omega/2 < \left|{\rm t} - \mu\right|$ and is zero otherwise.  The topological phase transition takes place at $|\mu|={\rm t}$, with the system being in the topological (trivial) phase for $|\mu| < {\rm t}$ ($|\mu| > {\rm t}$).  In Fig.~\ref{topo_trans} we plot $\Pi''(\omega)$ (main plot) and $\Pi'(\omega)$ (inset) as a function of the chemical potential $\mu$ for various values of the cavity frequency $\omega$. We see that this function shows a large peak at the transition point ($|\mu|={\rm t}$), which becomes narrower and more pronounced  for smaller  $\omega$ (compared to the gap $\Delta$). Physically, this is due to the fact that the electronic levels close to the zero energy  have larger curvatures, i.e. they are more susceptible  close to the phase transition point. The real part  also serves for detecting the phase transition, although not as directly as the imaginary part, as shown in Fig.~\ref{topo_trans}, where the phase transitions are inferred from the kinks in this function.  We have checked that the same peak structure holds for the cases when $\Delta\neq {\rm t}$,  too, the only modification being a shift in the scale for $\omega$, which should be of the order of $\omega\sim\Delta$.

\subsection{Spin-orbit (SO) coupled nanowire}

\begin{figure}[t] 
\includegraphics[width=0.95\linewidth]{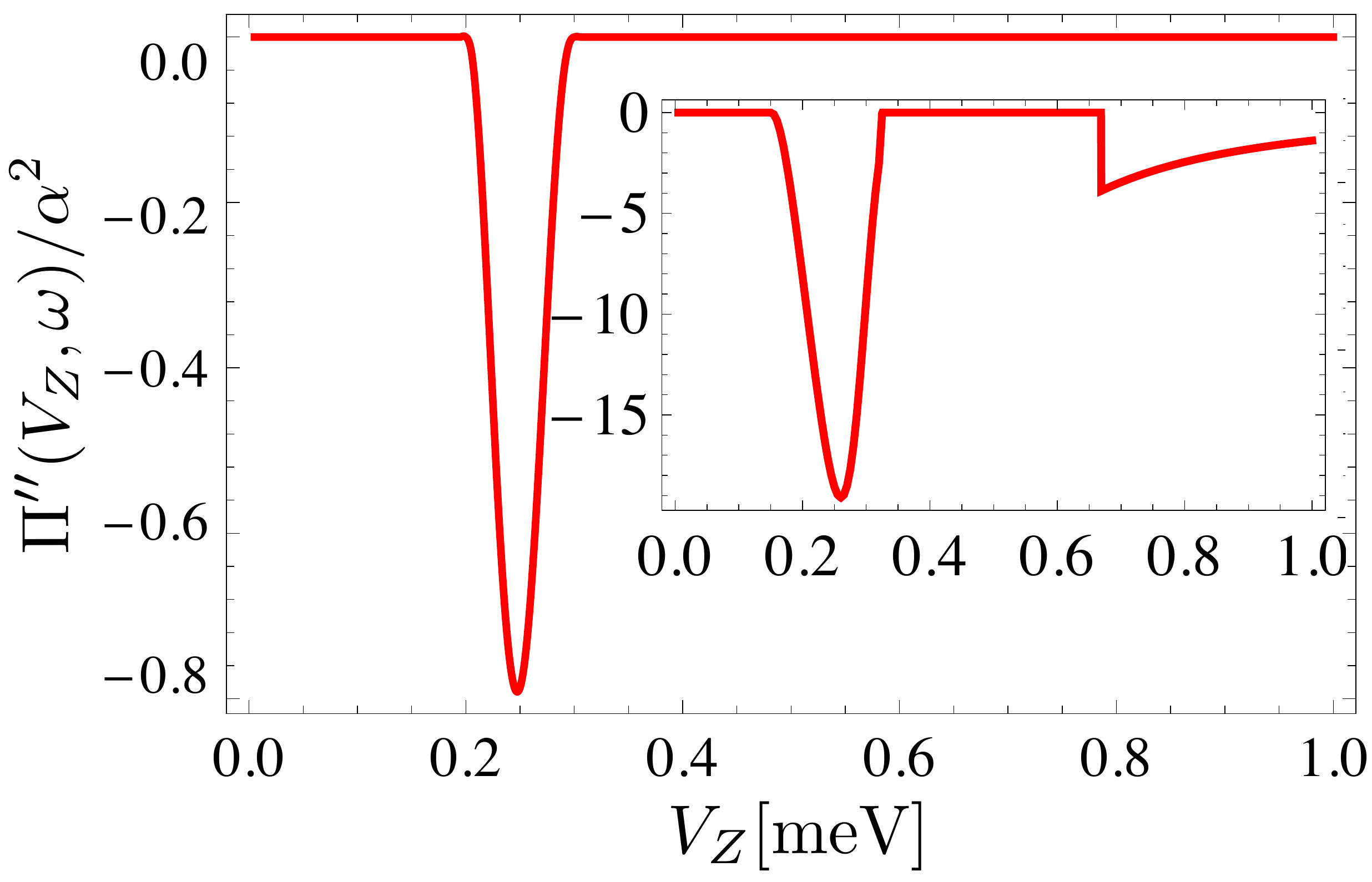}
\caption{The imaginary part of the density-density correlation function [$\Pi''(\omega)$] as a function of  the Zeeman splitting $V_Z$ for a SO coupled nanowire  in case of PBCs for parameters  $L=2\,\mu$\text{m}, ${\rm t}=0.5\cdot 10^{-2}$ \text{eV},  $\alpha=0.4$ \text{meV}, $\mu=-10^{-2}$ \text{eV}, $\Delta=0.25$ \text{meV}, $N=80$, and  $\omega=0.1$ meV ($\omega=0.2$ meV in the Inset). The topological transition takes place for $V_Z\approx 0.25$ meV, for which the susceptibility reaches its maximum. The emergence of a second peak is due to the resonance condition around the external gaps ($k\sim k_F$). } 
\label{topo_trans_SO} 
\end{figure}

The case of a realistic SO coupled nanowire is more complicated that the Kitaev model showed above, and so is the evaluation of susceptibility. This is so because the SO coupled wire has four bands (because of the spin), instead of two, and a more complicated quasiparticle spectrum. Nevertheless, writing the electronic operators in the Fourier space as $c_{j\sigma}=1/\sqrt{N}\sum_{k}e^{ikj}c_{k\sigma}$, we can write again the electronic Hamiltonian as $H_{el}^W=\sum_{k}H_{BdG}^W(k)$, with
\begin{equation}
H_{BdG}^W=[(-{\rm t}\cos{k}-\mu)+\gamma\sin{k}\,\sigma_z]\tau_z^k+V_Z\sigma_x+\Delta\tau_x^k\,,
\end{equation}
and the coupling to the cavity the same as in Eq.~\eqref{ElCav}. However, the expression for $\Pi(t)$ becomes rather cumbersome for the general case and  to get some analytical insights we need to resort to approximations.  For that, the  Hamiltonian can be put in a different form by the use of a unitary transformation (see Appendix~\ref{nanowire_periodic}):
\begin{align}
H_{BdG}^W(k)&=\left[-t\cos{k}-\mu+\sqrt{(\gamma\sin{k})^2+V_Z^2}\,\sigma_z\right]\tau_z^k\nonumber\\
&+\frac{\Delta\gamma\sin{k}}{\sqrt{(\gamma\sin{k})^2+V_Z^2}}\tau_x^k-\frac{\Delta V_Z}{\sqrt{(\gamma\sin{k})^2+V_Z^2}}\sigma_y\tau_y^k\,,
\end{align} 
while the $H_{el-c}$ stays unchanged. Progress can be made if we assume the limit of large magnetic field, $V_{Z}\gg\Delta,\mu$, in which case we can neglect the last term in the above Hamiltonian. By doing so, we recover two copies of the Kitaev chain, for $\sigma_z=\uparrow,\downarrow$. The susceptibility becomes:
\begin{equation}
\Pi(\omega)=-\alpha^2\sum_{k > 0;p,\sigma=\pm} \frac{\displaystyle \left(\Delta_{\rm eff}\sin k\right)^2}{\displaystyle E_{k\sigma}^2}\frac{\displaystyle p}{\displaystyle \omega + 2pE_{k\sigma} + i\eta}\,,
\end{equation}
where $E_{k\sigma}$ are given by the Kitaev spectrum with:
\begin{align}
\mu^{\rm eff}_{k\sigma}&=\mu-\sqrt{(\gamma\sin{k})^2+V_Z^2}\,\sigma\,,\\
\Delta_{\rm eff}&=\frac{\Delta\gamma}{\sqrt{(\gamma\sin{k})^2+V_Z^2}}\,.
\end{align}
All the results from the previous section apply to this case but with the $k$-dependent parameters showed above. The system is in the topological nontrivial (trivial) regime for $V_Z>\sqrt{\mu^2+\Delta^2}$ ($V_Z<\sqrt{\mu^2+\Delta^2}$). In Fig.~\ref{topo_trans_SO} we plot the imaginary part of the susceptibility as a function of the Zeeman splitting $V_Z$ for two different values of the cavity frequency $\omega$. We see a  similar behavior as in the case of the Kitaev chain: a peak emerges in $\Pi''(\omega)$ at the topological phase transition point, which becomes narrower as omega becomes smaller. However, an extra peak emerges at a larger  $V_Z$, and it is due to the resonance condition with the gaps around the $k\sim k_F$ (external gaps in the SO coupled nanowire spectrum). 

\section{Majorana fermions detection}\label{sec4}

In this section, we consider a finite 1D topological system coupled to the cavity (therefore with open boundary conditions, or OBCs), so that there are two Majorana fermions  emerging in the topological region, each localized at one of the two ends of the chain. Taken together, they give rise to a zero-energy fermionic state in the infinite wire limit, which can be either empty or occupied, thus labeling the {\it parity} of a 1D $p$-wave superconductor \cite{alicea2011non}. The Majorana wavefunctions  decay exponentially in the wire on the scale of the superconducting correlation length $\xi$, and for a finite wire it can lead to a finite energy splitting $\epsilon_M\propto\exp{(-L/\xi)}$ of the initially zero energy fermionic state~\cite{kitaev2001unpaired}. In the following, we will show that both the presence of the Majorana fermions and the parity of the Majorana fermionic state can be inferred from the susceptibility $\Pi(\omega)$. 

In the finite chain case we cannot obtain exact results for $\Pi(\omega)$ anymore, therefore we proceed to calculate this quantity numerically (see Appendix~\ref{openboundaries}). We will treat the two models, the Kitaev chain and the SO coupled wire on equal footing, showing that they give similar results. 

For starters, the electronic Hamiltonian can be casted in the following form:
\begin{equation}
H_{el}=\frac{1}{2}\vec{c}^\dagger M\vec{c}\,,
\end{equation}
with 
\begin{equation}
\vec{c}=(\{c_{1s}\}\dots\{c_{Ns}\}\{c^\dagger_{1s}\}\dots\{c^\dagger_{Ns}\})\,,
\end{equation}
where $s$ counts internal degrees of freedom, such as spin, band index, etc. For the Kitaev chain $s=1$ (and thus we can disregard it), while for the SO coupled nanowire $s=\uparrow,\downarrow$.  Here,  $M$ is a $2Ns\times 2Ns$ matrix~\cite{kitaev2001unpaired}, and   we can write $M=PWP^T$,  with
\begin{equation}
 W_{2p-s,2k-s}=(-1)^{s+1}\delta_{p,k}\epsilon_{k};~~s=0,1.
 \end{equation}
 $P$ is a unitary matrix ($PP^T=P^T P=1$) whose columns are the eigenvectors of $M$~\cite{kitaev2001unpaired}. Also,  $\epsilon_p$, with $p=1,\dots, sN$ are the eigenenergies of the electronic Hamiltonian, including the Majoranas (if present).  Thus, the electronic Hamiltonian can be re-written as 
$$H_{el}=(1/2)\vec{\widetilde C}^\dagger W\vec{\widetilde C},$$
 and $\vec{\widetilde{C}}=P^\dagger\vec{C}$, where 
\begin{equation}
\vec{\widetilde{C}}=(\{\tilde{c}_{1s}\}\dots\{\tilde{c}_{Ns}\}\{\tilde{c}^\dagger_{1s}\}\dots\{\tilde{c}^\dagger_{Ns}\})\,,
\end{equation}
with $\tilde{c}_{p}^\dagger$ ($\tilde{c}_p$) are the creation (annihilation) operators for the Bogoliubov quasiparticles in the finite wire, with $p=1\dots N$ labeling the energy levels. Finally, we can write 
\begin{equation}
H_{el}=\sum_{p,s}\epsilon_{ps}\left(\tilde{c}_{ps}^\dagger\tilde{c}_{ps}-\frac{1}{2}\right)\,,
\end{equation}
and also define the spinorial wavefunction for the state of energy $\pm\epsilon_{ps}$ at position $j$ as  $\vec{\psi}_{ps}(j)=(u^j_{ps},v^j_{ps})^T$, where $u^j_{ps} (v^j_{ps})=P_{2j-1,p}(P_{2j,p})$ are the electron (hole) components of the wavefunction at position $j$ in the wire. 

The  electron-cavity coupling Hamiltonian can be then written in the new basis as follows:
\begin{align}
H_{el-c}&=\sum_{p,p'}\left[C_{ps,p's'}^{(1)}\tilde{c}_{ps}^\dagger\tilde{c}_{p's'}-iC_{ps,p's'}^{(2)}\tilde{c}_{ps}^\dagger\tilde{c}_{p's'}^\dagger+{\rm h. c.}\right]\nonumber\\
&\times(a^\dagger+a)\,,
\label{el_cav}
\end{align} 
where  $C_{ps,p's'}^{(1,2)}$ are coefficients that depend on the transformation from the electronic basis $\vec{C}$ to the Bogoliubov basis $\vec{\tilde{C}}$ and read \cite{trif2012resonantly, cottet2015electron}:
\begin{align}
C_{ps,p's'}^{(1,2)}=\alpha\sum_{j=1}^N\vec{\psi}_{ps}^\dagger(j)\tau_{z,y}\vec{\psi}_{p's'}(j)\,.
\end{align}
Here the pseudo-spin $\vec{\tau}=(\tau_x,\tau_y,\tau_z)$ acts in the Nambu (or particle-hole) subspace. In general, all $C_{ps,p's'}^{(1,2)}\neq0$ for $p\neq p'$ and $s\neq s'$, thus there are couplings between all the levels (and bands) via the cavity field, and that includes transitions between the Majorana and the bulk (or gaped) modes. This in turn affects the correlation function in Eq.~\eqref{susc_time}, which can be written as $\Pi(\omega) = \Pi_{BB}(\omega) + \Pi_{BM}(\omega) + \Pi_{MM}(\omega)$, being the sum of the terms that contain only bulk states (bulk-bulk, or BB), cross terms between Majorana and the bulk (bulk-Majorana or BM), and Majorana contributions only (Majorana-Majorana or MM), respectively. However, $\Pi_{MM}(\omega)\equiv 0$~\cite{cottet2013squeezing} due to the fact that the cavity cannot mix different  parities, and in consequence the only contribution from the Majorana modes comes through the cross terms $\Pi_{BM}(\omega)$. We have found that  for $N\gg1$ the $\Pi_{BB}(\omega)$ contribution is given by the one obtained from the PBCs in the first part of the paper, i.e., $\Pi_{BB}\propto N$, while $\Pi_{BM}\propto const$, up to exponentially small terms in $L/\xi$. We note in passing that in a real wire, the smallness of the $\Pi_{BM}$ compared to $\Pi_{BB}$ is measured by  $\lambda_F/L$, with $\lambda_F$ being the Fermi wavelength and $L$ the length of the wire.  

\begin{figure}[t] 
\centering
\includegraphics[width=0.95\linewidth]{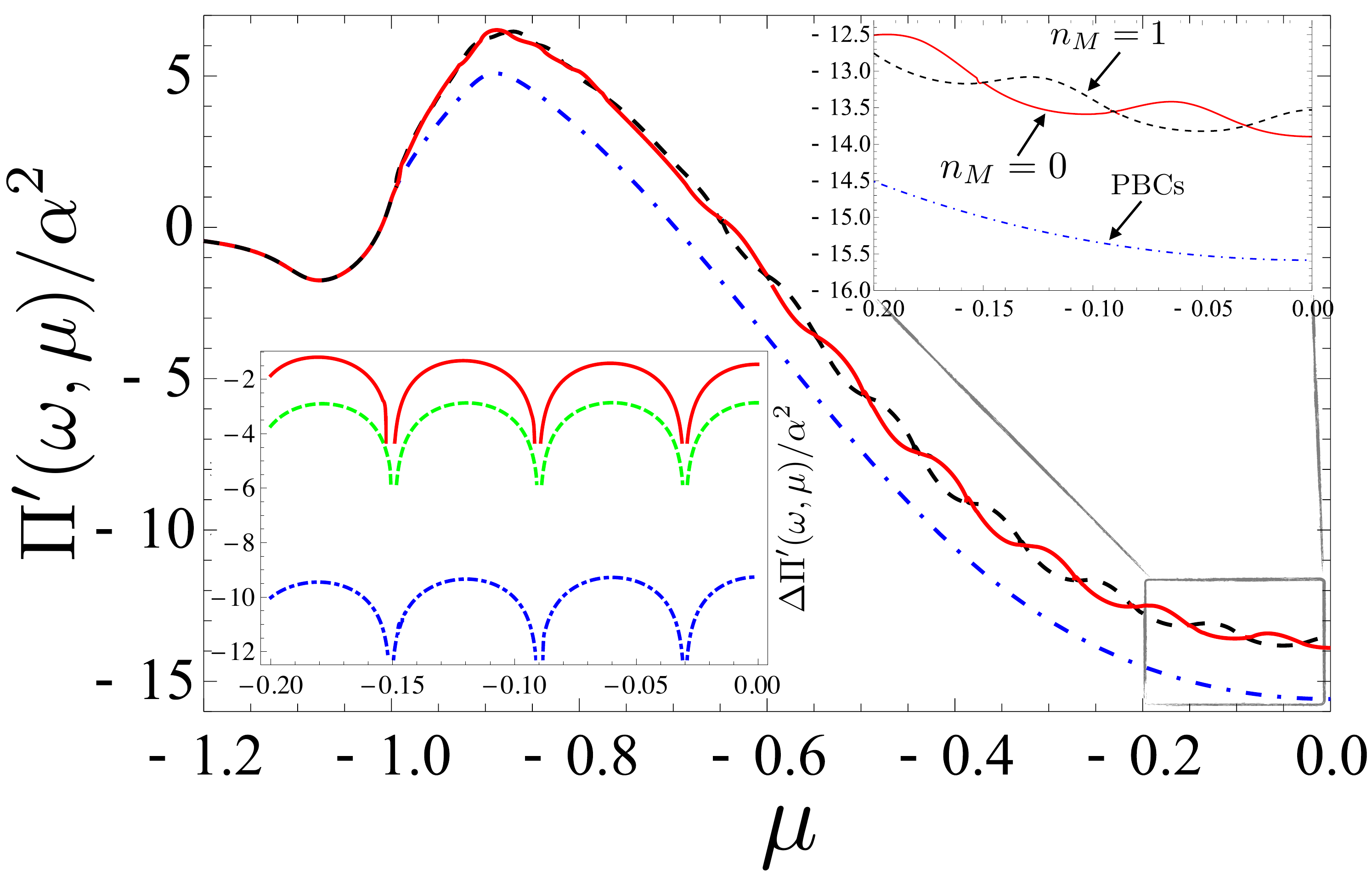}
\caption{Dependence of $\Pi'(\omega,\mu)$ on the chemical potential $\mu$. The  blue (dot-dashed), red (full), and the black (dashed) lines correspond to the susceptibility for PBCs,  OBCs for parity $n_M=0$, and OBCs for parity $n_M=1$, respectively. 
Lower Inset:  the relative strength of the susceptibility $\Delta\Pi'=2\left|(\Pi_{BM}^+-\Pi_{BM}^-)/(\Pi_{BM}^++\Pi_{BM}^-)\right|$  as a function of $\mu$ in logarithmic scale, for $\Delta=0.1$ (red-full), $\Delta=0.2$ (green-dashed), and $\Delta=0.3$ (blue-dot-dashed). The size of $\Delta\Pi'$ is exponentially reduced as a function of $\Delta$. We used  $N=50$,  $\omega=0.2$, $\Delta=0.1$, ${\rm t}=1$, and all energies are expressed in terms of ${\rm t}$. } 
\label{parities} 
\end{figure}

In the following, we analyze the cross-terms contribution $\Pi_{BM}(\omega)$.  For $\epsilon_{M}\ll\epsilon_{p}\pm\omega$, with $p\neq M$, we obtain:  
\begin{align}
&\Pi_{BM}(\omega)=\sum_{p,s\neq M}\left(\frac{1}{\epsilon_{ps}+\omega+i\eta}+\frac{1}{\epsilon_{ps}-\omega-i\eta}\right)\nonumber\\
\!\!\!&\times\left[|C_{M,ps}^{(1)}|^2(n_M-n_{ps})-|C_{M,ps}^{(2)}|^2(n_M-1+n_{ps})\right]\,,
\label{pibm}
\end{align}
where $n_{ps}$ and $n_M$ are the occupations of the bulk and Majorana states, respectively. This is one of our main results. Inspecting the above expression, we see that it is strongly dependent on the Majorana state parity $n_M$. Assuming that $\epsilon_{ps}>0$ for $p,s\neq M$ and $n_{ps}=0$ for $n\neq M$ in the ground state, we obtain that $\Pi_{BM}^+\propto |C_{M,ps}^{(1)}|^2$ ($\Pi_{BM}^-\propto |C_{M,ps}^{(2)}|^2$) for $n_M=1$ ($n_M=0$). To get more physical insight into the resulting susceptibility, we write the coefficients $C_{M,ps}^{(1,2)}$ in the following way:
\begin{align}
C_{M,ps}^{(r)}&=\sum_j[(u_M^j\delta_{r,1}+v_M^j\delta_{r,2})u_{ps}^j\nonumber\\
&-(u_M^j\delta_{r,2}+v_M^j\delta_{r,1})v_{ps}^j]\,.
\end{align}
Let us analyze the implication of the above result. When $\epsilon_M=0$, we also have $u_M^j=v_M^j$, and thus $C_{Mp}^{(1)}=C_{Mp}^{(2)}$, since electron and hole contributions are the same in the Majorana state. However, for a finite energy splitting $\epsilon_M\neq 0$, and thus we have that $u_M^j\neq v_M^j$, which in turn results in  $C_{Mp}^{(1)}\neq C_{Mp}^{(2)}$. All these suggest that the susceptibility  $\Pi(\omega)$, via $\Pi_{BM}(\omega)$ should allow us to infer both the parity of the ground state and the zeros in the Majorana energy $\epsilon_M$, assuming their spatial overlap is large enough. 
 
\begin{figure}[t] 
\centering
\includegraphics[width=0.96\linewidth]{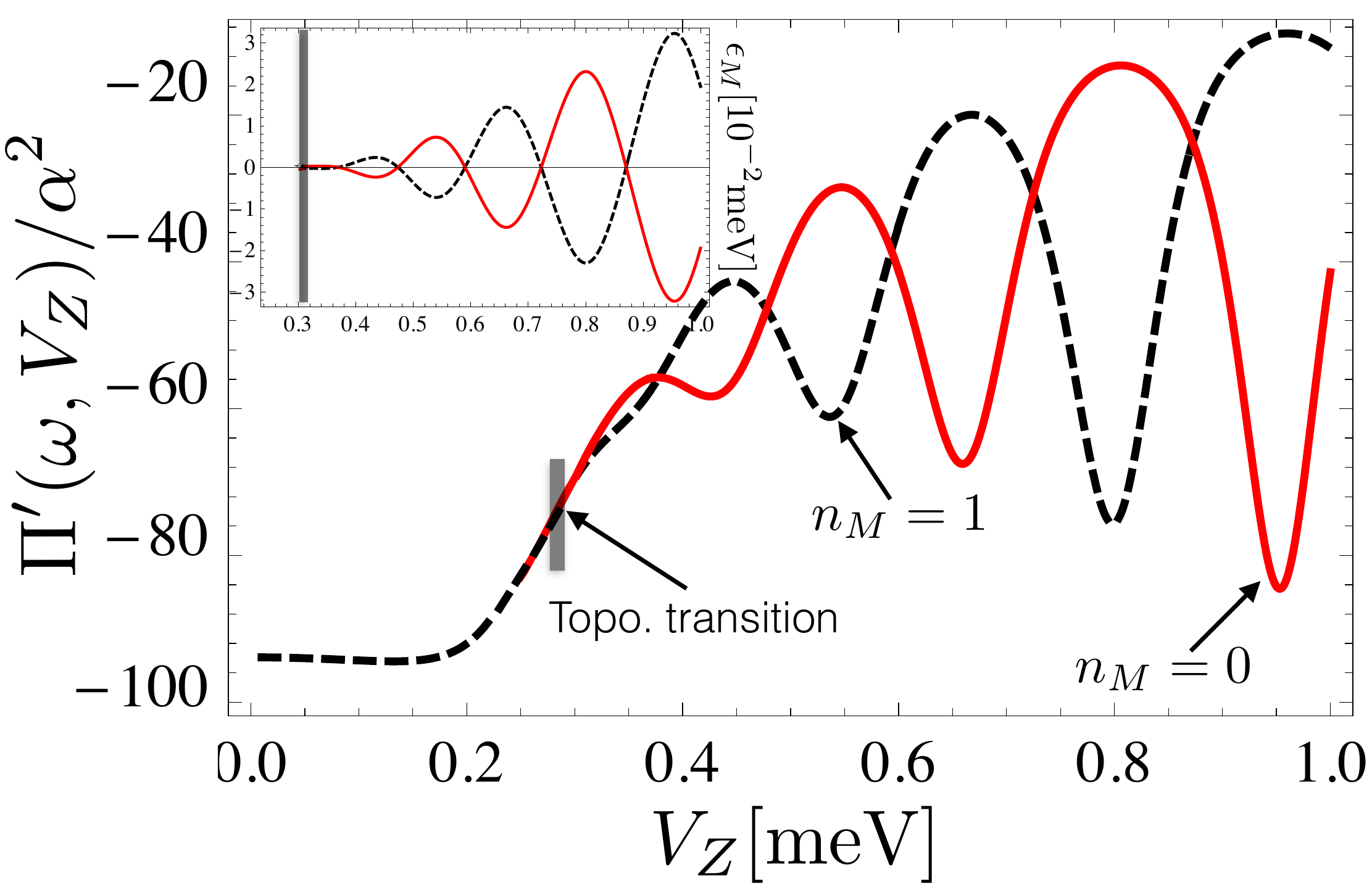}
\caption{Dependence of $\Pi'(\omega,V_Z)$ on the Zeeman splitting $V_Z$. Above the topological transition (indicated in the figure), the red (full) and black (dashed) curves correspond to $\Pi'(\omega,V_Z)$ for parity $n_{M}=0$ and $n_M=1$, respectively. $\Pi'(\omega,V_Z)$ shows oscillations as a function of $V_Z$, that are different in amplitude for the two parities $n_M=0,1$ (but having the same period), crossing at points where $\epsilon_M=0$ (see Inset). Below the topological transition the susceptibility reproduces well the one for PBCs. The parameters we took  are \cite{Halperin2015}: $L=2\mu$\text{m}, $t=0.5\cdot 10^{-2}$\text{eV},  $\alpha=0.4m$\text{eV}, $\mu=-10^{-2}$\text{eV}, $\Delta=0.25m$\text{eV},  $\omega=0.02m$\text{eV}, and $N=80$. } 
\label{parities-wire} 
\end{figure}

In the main plot in Fig.~\ref{parities}, we show the real part $\Pi'(\omega)$ for the Kitaev chain  as a function of the chemical potential $\mu$ for the two parities $n_{M}=0,1$ as well as the bulk value for PBCs. 
First of all, the values for $\Pi(\omega)$ in case of periodic and OBCs are different because of $\Pi_{BM}(\omega)$, as this contribution  has a different dependence on $\mu$ and $\Delta$ from the bulk states. Second of all, the open BCs wire susceptibility shows oscillations as a function of $\mu$ on top of the average value, of the form $\pm\cos{(k_FL)}$,  with $+(-)$ corresponding to $n_M=1$ ($n_M=0$), i.e. they are opposite in sign for the two parities. Here $k_F$ is the Fermi wavevector of the electronic system, and  for the range of parameters considered is $k_F\approx2\mu$~\cite{kitaev2001unpaired}. This  means that the cavity field can access the parity of the Majorana fermions non-invasively and without locally accessing the wire. Moreover, the oscillations disappear below the phase transition point $|\mu|=1$, the susceptibility $\Pi'(\omega)$ acquires the same value as for the PBCs wire which signals  that the Majorana fermions exist only above the topological phase transition.  In order to get a closer look at the oscillations of $\Pi(\omega,\mu)$,  in the lower inset  in Fig.~\ref{parities} we show the real part of the relative difference between the two parities, $\Delta\Pi(\omega,\mu)=2\left|(\Pi_{BM}^{+}-\Pi_{BM}^{-})/(\Pi_{BM}^{+}+\Pi_{BM}^{-})\right|$,  for different values of  $\Delta$. We see that the oscillations have the same periodicity as the Majorana energy splitting $\epsilon_M\sim\exp{(-L/\xi)}|\cos{(k_F L)}|$.
Notice that the  oscillations of the Majorana splitting with the chemical potential has been studied in detail 
\cite{prada2012oscillations,dassarma2012oscillations}  together with the fact that the magnitude of the oscillations becomes exponentially suppressed in $L/\xi$~\cite{pientka2013magneto, zyuzin2013correlations}. 

In the main figure in Fig.~\ref{parities-wire}, we plot the real part of the 
susceptibility for a 1D topological wire as a function of the Zeeman splitting 
$V_Z$ for the two parities  $n_{M}=0,1$.
The susceptibility $\Pi$ for that figure  was computed using realistic 
parameters that might be appropriate
for an InSb wire such as in the experiments in 
Ref.~\onlinecite{mourik2012signatures}. We find similar features as for the Kitaev toy model, namely oscillations as a function of the Zeeman splitting above
the topological transition. These oscillations around the ground state have opposite sign and different amplitudes for each parity. Like for the Kitaev model, they have the same periodicity as the Majorana energy $\epsilon_M$ (see  the inset of Fig.~\ref{parities-wire}) and cross at points where $\epsilon_M=0$. Notice that if the parity is not conserved in the system (for example, due to the quasi-particle poisoning), $\Pi'$ will follow the ground state and exhibit therefore sharp cusps as a function 
$\Delta_Z$ at the crossing points where $\epsilon_M=0$ (see also Ref.~\onlinecite{Vayrynen:2015} for similar features in a topological Josephson junction). As for the Kitaev chain, we thus find that the cavity phase shift is thus able to detect the Majorana fermions and the parity of the  ground state of a realistic topological wire.

\begin{figure}[t] 
\centering
\includegraphics[width=0.95\linewidth]{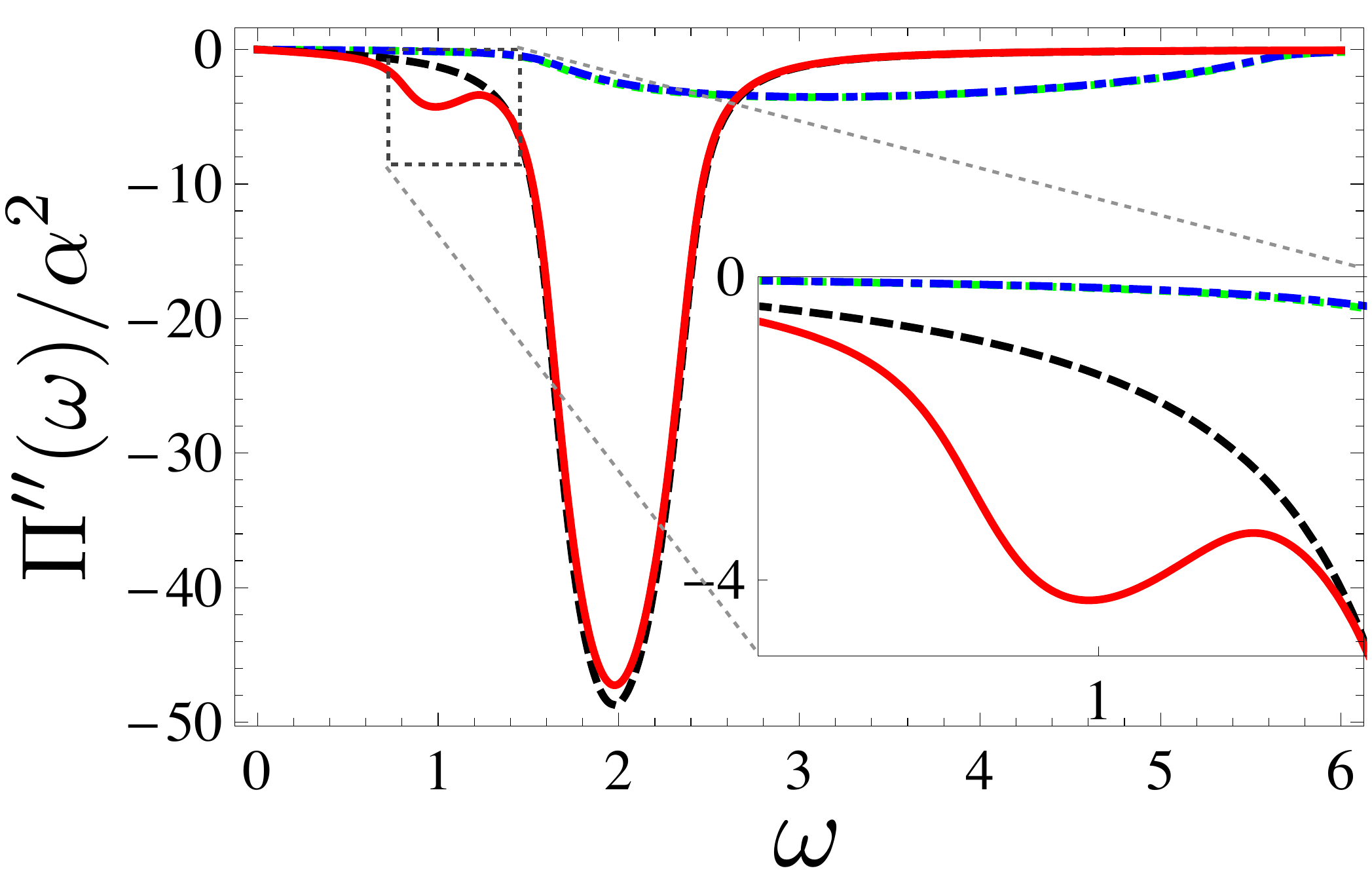}
\caption{Dependence of $\Pi''(\omega)$ on the cavity frequency $\omega$ for $N=50$. The full-red (dashed-black) line corresponds to the susceptibility in the topological regime for $\mu=-0.2$, while the dot-dashed-blue (dotted-green) corresponds to the non-topological regime with open (periodic) BCs with  $\mu=-1.8$, so that the effective gap is the same $\Delta_{\rm eff}=||\mu|-{\rm t}|=0.8$ in both regions. 
 Inset: a zoom in the region where the Majorana peak emerges. For all the plots we used  ${\rm t} = \Delta = 1$, and all energies are expressed in terms of ${\rm t}$.} 
\label{susc_of_omega} 
\end{figure}

The imaginary part of $\Pi(\omega)$ gives us also information on the presence of Majorana fermions. In Fig.~\ref{susc_of_omega}, we show the dependence of $\Pi''(\omega)$ on $\omega$ for the Kitaev chain, both in the topological and non-topological regimes, for ${\rm t}=\Delta$. We see that the Majorana fermions, through $\Pi_{BM}(\omega)$, give rise to an extra peak in the susceptibility at half the effective superconducting gap $\Delta_{\rm eff}=||\mu|-{\rm t}|$ in the topological regime, while such a peak is absent for the same effective gap $\Delta_{\rm eff}$, but in the non-topological case. For completeness, we also show the result for PBCs, in which case there are no Majorana fermions. 

Finally, let us give some estimates for $\Pi(\omega)$, and in particular for  $\Pi_{BM}(\omega)$ and the resulting phase shift in the exiting photonic signal. We assume  typical experimental values for the cavity frequency,   $\omega_c\approx2\times10^{-5}$ eV, and with a quality factor $Q\approx10^5$, which results in photon escape rate $\kappa=2\times10^{-10}$ eV. For an estimate of the capacitive coupling $\alpha$ we refer, for example, to  the case of carbon nanotubes, which have been under experimental scrutiny in the context of cavity QED \cite{cottet2015electron,delbecq2011coupling}.  There, it was found that $\alpha\approx5.6\times10^{-7}$ eV, and we believe similar values should be relevant for semiconductor nanowires too. 
The phase shift of the radiation exiting the cavity satisfies $\delta\phi\propto(\alpha^2/{\rm t}\,\kappa)$ so that we obtain $\delta\phi\approx 0.3$ which is a sizeable value.

\section{Conclusions and outlook}\label{sec5}

 We  studied two paradigmatic examples of 1D topological superconducting systems capacitively coupled to a microwave superconducting stripline cavity: the Kitaev chain and a 1D nanowire with strong SO interaction in the presence of a magnetic field and in proximity of a superconductor. We analyzed the electronic charge susceptibility of these systems that is revealed in the photonic transport through the microwave cavity via its transmission ${\bf \tau}(\omega)$. We showed that this electronic susceptibility can actually be used to detect the topological phase transition, the occurrence of Majorana fermions and  the parity of the Majorana fermionic state in a {\it non-invasive} fashion.  Such effects are due to the interplay between the bulk and Majorana states, either via virtual or real transitions taking place between  the two, and  which are mediated by the photonic field. As an outlook, it would be interesting to use the same cavity QED setup  to access the physics associated with the fractional Josephson effect.

\begin{acknowledgments}
{\it Acknowledgments}|
 We acknowledge discussions with the LPS mesoscopic group in Orsay, and in particular insightful  suggestions by  Helene Bouchiat,  and we thank Jukka Vayrynen for valuable correspondence. This work is supported
by a public grant from the ``Laboratoire d'Excellence Physics Atom Light Matter'' (LabEx PALM, reference: ANR-10-LABX-0039) and the French Agence
Nationale de la Recherche through the ANR contract Dymesys.
\end{acknowledgments}

\appendix

\section{Derivation of the effective Kitaev Hamiltonian in the presence of the cavity field} 
\label{sec:effective}

In this section, we provide theoretical arguments for the wire Hamiltonian utilized  in Eq.~(\ref{kitaevhamiltonian}), and the effective electron-cavity Hamiltonian used   in the Main Text (MT). In a continuum description, the natural way to account for the interaction between the electrons and the electromagnetic field is via the minimal coupling, i.e. ${\bf p} \rightarrow {\bf p}-(e/c){\bf A}$ in the electronic Hamiltonian, with ${\bf A}$ being the electromagnetic field vector potential and ${\bf p}$ being the momentum of the electrons in the material.  In a tight-binding picture instead, one accounts for the coupling between light and matter by performing the Peierls substitution to the hopping parameters $t_{ii+1}$ between neighboring sites $i$ and $i+1$, namely 
\begin{equation}
t_{ii+1}\rightarrow t_{ii+1}e^{i\int_{i}^{i+1}{\bf A}({\bf r})\cdot d{\bf r}}\,,
\end{equation}
with ${\bf A}({\bf r})$  being the electromagnetic field vector potential at position ${\bf r}$, and the integration is performed between the sites $i$ and $i+1$.  We will focus on the derivation of the effective Kitaev model in the tight-binding picture, as the microscopic, continuum model was described in great detail very recently in~\cite{cottet2015electron}. We thus refer the reader to that paper for a detailed calculation of the cavity effects, as well as the derivation of the capacitive coupling starting from the minimal coupling.   

Here we give some details on the derivation of Eq.~(\ref{kitaevhamiltonian}) in the MT starting from a non-superconducting nanowire coupled to a bulk $p$-wave superconductor with such a coupling being assisted by the cavity field. For simplicity, we assume the bulk to be not $s$, but $p$-wave paired, thus the presence of spin-orbit coupling  in the wire is not a necessary ingredient. However, the present calculations can be straightforwardly generalized to more realistic system, such as nanowires with SOI.  The total  Hamiltonian of the system reads: 
\begin{align}
H_{\rm sys}&= H_b+H_w+ H_T+ H_c\,,
\label{h0}
\end{align}
where 
\begin{align}
H_{p} &=-\mu_{p}\sum_jc_{j,p}^\dagger c_{j,p}+\sum_{j} \Big(t_p c_{j,p}^\dagger c_{j+1,p}\nonumber\\
&+\Delta_p c^\dagger_{j,p}c^\dagger_{j+1,p}+{\rm h.c.}\Big)
\end{align}
with $p=b(bulk),w(wire)$, and $\Delta_w=0$ (no intrinsic superconductivity in the wire), and $\Delta_b\equiv\Delta$ the $p$-wave pairing in the bulk superconductor. Here, $c_{j,p}$ ($c_{j,p}^\dag$) and  $t_p$ are the electronic annihilation (creation) operator at position $j$ and the hopping parameter in system $p=b,w$, respectively. The tunneling Hamiltonian in the presence of the cavity reads:
\begin{align}
H_T = \sum_{j}\left(t_{int}e^{-i\hat{\phi}_j}c^\dag_{j,w}c_{j,b} +{\rm h.c.}\right),
\label{htunneling}
\end{align}
where  $\hat{\phi}_j=\hat{A}_{j}d_j$, with  $\hat{A}_j=i(\alpha_j/\omega_c)(a^\dagger-a)$, $d_j$, $\alpha_j$, $\omega_c$, and  $a$ ($a^\dagger$),  being the cavity vector potential, the  coupling strength, the cavity frequency, and the cavity photon annihilation (creation) operators, respectively. Note that we assumed that the cavity field points perpendicularly to the wire, and it has no component along it. If instead such components would exists, we should have modified the wire Hamiltonian too in order to account for the cavity induced phase factors. In the following, we will assume that $\alpha_jd_j\equiv \alpha_j=\alpha$, namely it is constant along  the entire wire. Finally, the Hamiltonian of the cavity reads:
\begin{align}
H_c = \omega_ca^\dag a\,,
\label{hcavity}
\end{align}
with $\omega_c$ being the (fundamental) frequency of the cavity. Before deriving an effective wire Hamiltonian, it is instructive to switch to the Fourier space, for both the bulk and wire Hamiltonians. We get:        
\begin{align}
H_b&=\sum_{k}\xi_{k,b}c^\dag_{k,b}c_{k,b}- \sum_{k>0}i\Delta\sin{k}\left(c_{-k,b}c_{k,b} - c^\dag_{k,b}c^\dag_{-k,b}\right)\,,\\
H_w&= \sum_{k}\xi_{k,w}c^\dag_{k,w}c_{k,w}\,,\\
H_{T}&= t_{int}\sum_{k}\left(e^{i\hat{\phi}}c^\dag_{k,w}c_{k,b} + e^{-i\hat{\phi}}c^\dag_{k,b}c_{k,w}\right),
\label{hwire}
\end{align}
where $\xi_{k,p}=t_p\cos{k}-\mu_p$, with $\mu_\alpha$ the chemical potential in the $p=w,b$ system. 

Next we perform the so called  Lang-Firsov transformation on the system Hamiltonian, which means  $\widetilde{H}_{sys} = \exp(S)H_{sys}\exp(-S)$ with  $S$ chosen as follows:
\begin{align}
S = \frac{\alpha}{\omega_c}(a - a^\dag)\sum_qc^\dag_{q,w} c_{q,w}.
\end{align}
After some lengthy, but straightforward calculation we obtain the system Hamiltonian as follows:

 

\begin{align}
\widetilde{H}_{sys} &= H_w+H_b+\underbrace{\alpha \sum_qc^\dag_{q,w} c_{q,w}(a + a^\dag)}_{H_{c-w}}\nonumber\\
&+\frac{\alpha^2}{\omega_c}\bigg(\underbrace{\sum_qc^\dag_{q,w} c_{q,w}}_{\hat{N}^2}\bigg)^2
+\underbrace{t_{int}\sum_{k}(c^\dag_{q,w}c_{k,b} + {\rm h.c.})}_{H_T}+ H_c \,,
\end{align}
which implies we excluded the photonic field from the tunneling term at the expense of adding photon-dependent chemical potential shift in the wire (third term) as well as an interaction term (fourth term). Note that for $t_{int}=0$, the transformation does not affect the spectrum, as it can be simply undone. However, as will see in the following, in the presence of the tunneling term the photonic field in the form of the capacitive coupling can lead to  real effects. 

In the following, we aim at finding an effective Hamiltonian describing the wire only by integrating the bulk superconductor degrees of freedom up to second order in the tunneling $t_{int}$. We choose to do so by employing the  
Schrieffer-Wolff transformation formalism, which means, as before, that we unitary rotate the system Hamiltonian as 
\begin{align}
H^{\rm eff}_{sys}&=e^{S_{SW}}\widetilde{H}_{sys}e^{-S_{SW}}=H_w+H_b+H_{w-c}+H_T+H_{c}\nonumber\\
&+[S_{SW},H_w+H_b+H_{w-c}+H_T+H_{c}]+\dots\,,
\end{align}
and choose 
\begin{align}
\left[S_{SW}, H_{w}+H_{b}\right] = -H_T\,,
\label{condition}
\end{align}
or $S_{SW}=(\mathcal{L}_w+\mathcal{L}_b)^{-1}H_T$, with $\mathcal{L}_{\alpha}$ being a superoperator whose action is defined as  $\mathcal{L}_\alpha A=[H_\alpha,A]$, $\forall A$. This is equivalent to the following identity:
\begin{align}
S_{SW}=i\lim_{\eta\rightarrow 0}\int_{0}^{+\infty}dte^{-\eta t}e^{i(H_w+H_b)t}H_Te^{-i(H_w+H_b)t}.
\label{s}
\end{align}
This term excludes the tunneling Hamiltonian $H_T$ in leading order (assuming there is no diagonal contribution caused by such a term). Then, we neglect the contributions of the higher order terms on the wire spectrum by averaging over the bulk ground state $|0_b\rangle$ in order to derive a purely (renormalized) wire Hamiltonian: 
\begin{align}
H^{\rm eff}_{w}&\approx\langle 0_b|H_{b} + H_{w} +H_{c}+ H_{w-c} +\frac{1}{2}\left[S_{SW}, H_T\right]\nonumber\\
&+ \left[S_{SW}, H_{w-c}\right]+\dots\,|0_b\rangle\,,
\label{htransform}
\end{align}
In order to find $S_{SW}$ from Eq.~{\eqref{s}} explicitly,  let us perform Bogoliubov transformation for the bulk $p$-wave superconductor defined as
\begin{align}
c_{k,b} = u_k^*\gamma_{k,b} + v_k\gamma^\dag_{-k,b}\,,\\
c^\dag_{-k,b} = -v_k^*\gamma_{k,b} + u_k\gamma^\dag_{-k,b}\,,
\end{align}
where $k>0$ and
\begin{align}
u_k&= \sqrt{1/2\left(1+\xi_k/E_{kb}\right)}\nonumber\\
v_k&= \sqrt{1/2\left(1-\xi_k/E_{kb}\right)}e^{-i\phi_b} 
\end{align}
with $\phi_b$ the phase of the superconducting condensate (that we choose $=0$ from now on) and $E_{kb} =\sqrt{\xi_k^2 + \Delta^2\sin^2{k}}$ the spectrum. 
We can then express the bulk Hamiltonian in terms of the $\gamma_{k}$ and $\gamma_{-k}$ operators:
\begin{align}
H_{b} = \sum_{k>0}E_{kb}\left(\gamma^\dag_{k,b}\gamma_{k,b} + \gamma^\dag_{-k,b}\gamma_{-k,b}\right)\,.
\label{hbulkb}
\end{align}
Utilizing the fact that:
\begin{align}
c_{k,w}(t) = c_{k,w}(0)\exp(-i\xi_{k,w}t)\,,
\end{align}
and
\begin{align}
\gamma_{k,b}(t) = \gamma_{k,b}(0)\exp(-iE_{k,b}t)\,,
\end{align}
we can readily find the transformation matrix $S_{SW}$ as follows (assuming also that $\xi_{k,w}\ll E_{k,b}$, since we are interested in the energies well inside the band gap of the bulk superconductor):
\begin{align}
S_{SW}&= \sum_k\frac{t_{int}}{E_{kb}}\Big[\left(|u_k|^2 - |v_k|^2\right)(c^\dag_{k,w}c_{k,b}-c^\dag_{k,b}c_{k,w})\nonumber\\
&-2u_kv_k(c^\dag_{k,w}c^\dag_{-k,b}-c_{-k,b}c_{k,w})\Big]\,.
\end{align}
Utilizing this expression for $S_{SW}$, we can calculate the expectation values for the different commutators in Eq.~\eqref{htransform}. We obtain:
\begin{align}
&H_{ind,w}\equiv \frac{1}{2}\langle0_b|\left[S_{SW}, H_T+2H_{c-w}\right]|0_b\rangle
\approx-\sum_{k}\frac{t_{int}^2}{E_{kb}}\nonumber\\
&\times\Big[\left(|u_k|^2 - |v_k|^2\right)c^\dag_{k,w}c_{k,w}-2u_kv_kc^\dag_{-k,w}c^\dag_{k,w}+{\rm h.c.}\Big]\,,
\end{align}
which can be interpreted as follows: the first term renormalizes the single particle spectrum in the wire, while the second term is responsible for the induced superconductivity in the wire. The full wire Hamiltonian thus becomes:
\begin{align}
H^{\rm eff}_{w}&=\sum_{k}(\underbrace{\xi_{k,w} + \delta\xi_{k,w}}_{\xi_{k,w}^{\rm eff}})c^\dag_{k,w}c_{k,w}\nonumber\\
&+ 2\sum_{k}\left(\Delta_{ind}c^\dag_{k,w}c^\dag_{-k,w} +{\rm h. c}\right)\nonumber\\
&+\alpha\sum_{k}c^\dag_{k,w}c_{k,w}\left(a + a^\dag\right)+\frac{\alpha^2}{\omega_c}\hat{N}_w^2
\label{eff_wire}
\end{align}
with 
\begin{align}
\delta\xi_{k,w}&=\frac{t_{int}^2}{E_{kb}}\left(|u_k|^2 - |v_k|^2\right)=\frac{t_{int}^2\xi_{k,b}}{E^2_{kb}}\,,\\
\Delta_{ind}&= \frac{t_{int}^2}{E_{kb}}u_kv_k=\frac{t_{int}^2}{2E_{kb}^2}\Delta\sin{k}
\end{align}
being the renormalization of the single-particle energies and the $p$-wave induced gap ($\propto\sin{k}$). Note that the last term in Eq.~\eqref{eff_wire} can be seen as a normalization of the single-particle spectrum in the mean-field, and thus finally we recover the same wire Hamiltonian defined in Eq.~(\ref{kitaevhamiltonian}) in the MT.


\section{Input-output theory for microwave cavities}\label{inputoutput}

\begin{figure}[t] 
\centering
\includegraphics[width=0.95\linewidth]{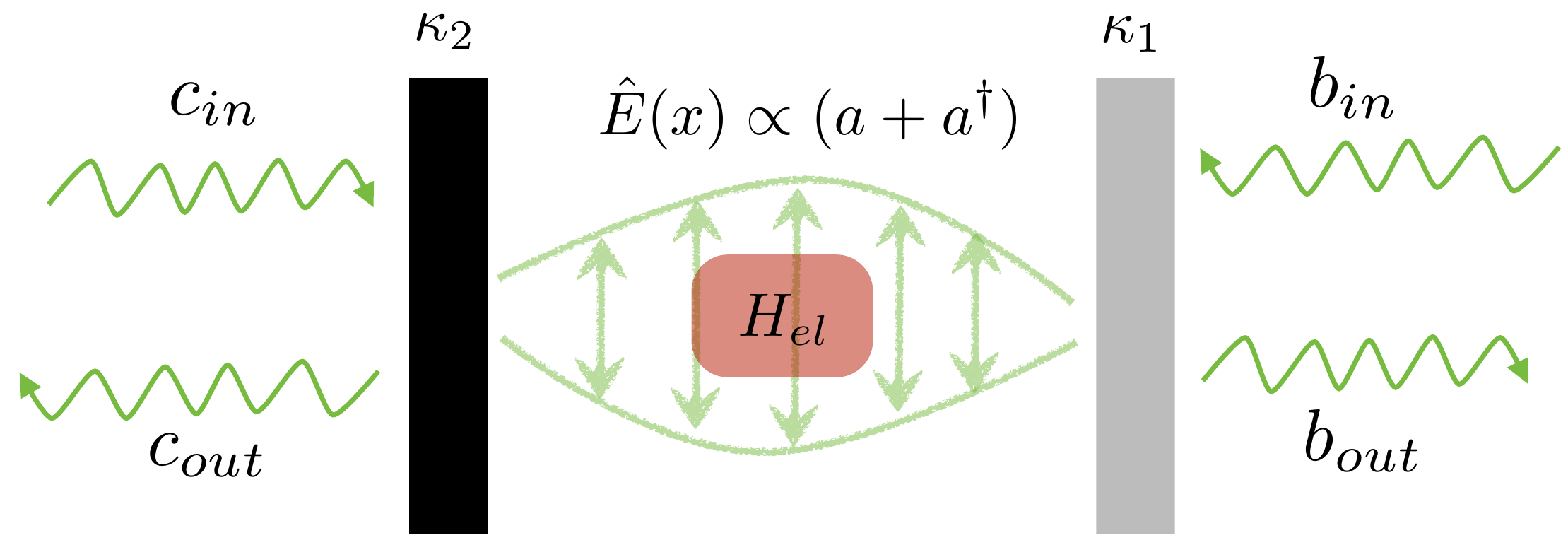}
\caption{Sketch of a one-sided cavity system probed in photonic transport. The input fields $b_{in}$ and $c_{in}$ are sent from the right and left mirrors, respectively, towards the cavity, and an output fields $b_{out}$ and $c_{out}$ are collected on the same sides. The cavity field, which is quantified by the bosonic operators $a$ and $a^\dagger$, interacts with the electronic system via capacitive coupling, affecting the cavity. The coupling between the cavity field and the external modes is quantified by the decay rate $\kappa_{1(2)}=2\pi\rho|f_{1(2)}|^2$, with $\rho$ the bath density of states and $|f_{1(2)}|^2\equiv f_{1(2)}f_{1(2)}^*$ being the average coupling between the cavity and the bath modes on the right (left). In the first part, for simplicity, we assume that $\kappa_2=0$, which correspond to a one-side cavity.} 
\label{sketch} 
\end{figure}

In this section, we present details on the input-output theory for the cavity in the presence of the coupling to a 1D $p$-wave SC. We will show that the transmission of the cavity depends on the electronic susceptibility of the electronic system. We concentrate on only single resonance of the cavity with frequency $\omega_c$~\cite{clerk2010introduction} and, for simplicity, we assume for the moment a single-sided cavity. The total Hamiltonian describing the system reads
\begin{align}
H =\underbrace{H_{el}+H_{c}+H_{el-c}}_{H_{sys}}+ H_{b} + H_{c-b}\,,
\end{align}
where $H_{el}$ is the electronic Hamiltonian only, and 
\begin{align}
H_{c}&=\omega_ca^\dagger a\,,\\
H_{el-c}&=\alpha(a+a^\dagger)n\,,\\
H_{b}&=\sum_{q}\hbar\omega_qb^{\dag}_qb_q\,,\\
H_{c-b}&=-i\hbar\sum_{q}\left(f_qa^{\dag}b_q - f^*_qb_q^{\dag}a\right)\,,
\end{align}
are the cavity, the electronic-cavity, the bath, and  the cavity-bath Hamiltonians, respectively.  Above, $a$ ($a^\dagger$) is the annihilation (creation) operator for the cavity mode with energy $\omega_c$, $b_{q}$ ($b_{q}^\dagger$) are the annihilation (creation) operators for the bath modes with energy  $\omega_q$, with $q$ labeling  their quantum numbers, and the complex coefficients  $f_{q}$ are coupling parameters between the cavity and the external bath. Moreover, $\alpha$ is the coupling strength between the cavity field and the total number operator in the system $n\equiv\sum_{j=1}^Nc_j^\dagger c_j$, with  $c_j$ ($c_j^\dagger$) being the annihilation (creation) operator for the electrons (fermionic degrees of freedom) at site $j$ in the electronic system.  

The idea of the input-output theory is to find the output photons (or field) in terms of the input ones, as shown schematically in Fig.~\ref{sketch}. 
 Following Ref.~\cite{clerk2010introduction}, we obtain for the cavity equation of motion for a one-side cavity ($\kappa_2=0$ in Fig.~\ref{sketch}):
\begin{align}
\dot{a} = \frac{\displaystyle i}{\displaystyle \hbar}\left[H_{sys}, a\right] - \frac{\displaystyle \kappa}{\displaystyle 2}a - \sqrt{\kappa}b_{in}\,,
\label{ain}
\end{align}
for the input field, and
\begin{align}
\dot{a} = \frac{\displaystyle i}{\displaystyle \hbar}\left[H_{sys}, a\right] + \frac{\displaystyle \kappa}{\displaystyle 2}a - \sqrt{\kappa}b_{out}\,,
\label{aout}
\end{align}
for the output field, where $\kappa=2\pi\rho|f|^2$ is the cavity decay rate, with $\rho$ the bath density of states and $f$ being the average coupling between the cavity and the bath modes. Subtracting Eq.~\eqref{aout} from Eq.~\eqref{ain} we obtain that
\begin{align}
b_{out}(t) = b_{in}(t) + \sqrt{\kappa}a(t), 
\label{input_output}
\end{align}
a result which holds  for any general cavity Hamiltonian. 

In the following, we will establish the relationship between $b_{out}$ and $b_{in}$ in the presence of the electronic system, as depicted in Fig.~\ref{sketch}.  For that, we first evaluate the commutator:
\begin{align}
[H_{sys},a]=-\omega_ca-\alpha n\,,
\label{eq_motion}
\end{align}
where $n$ is the time-dependent electronic particle number (see below). In order to utilize this contribution to the  equation of motion of the cavity field, we need to evaluate the time-dependent particle number $n(t)$, which itself depends on the coupling to the cavity. At time $t$, we can write:
\begin{align}
n_H(t) = e^{iH_{sys}(t - t_0)}ne^{-iH_{sys}(t - t_0)},
\end{align}
and 
\begin{align}
n_I(t) = e^{iH_{el}(t - t_0)}ne^{-iH_{el}(t - t_0)},
\end{align}
being the Heisenberg and interaction pictures, respectively, with $t>t_0$ being the initial time and which can be chosen at will. We can represent $n_H(t)$ as follows:
\begin{align}
n_H(t) = U^\dag(t, t_0)n_I(t)U(t, t_0),
\label{nheisenberg}
\end{align}
where
\begin{align}
U(t, t_0) = T_c\exp\left(-i\int_{t_0}^{t}dt'H_{el-c}(t')\right)
\end{align}
is the evolution operator with $T_c$ the time-ordering operator that puts operators with later times to the left of  the ones with earlier times. We can then write~Eq.(\ref{nheisenberg}) in the following way
\begin{widetext}
\begin{align}
&n_H(t) = T_c\exp\left(i\int_{t_0}^{t}dt'H_{el-c}(t')\right)n_I(t)T_c\exp\left(-i\int_{t_0}^{t}dt'H_{el-c}(t')\right)\nonumber\\
&\approx  \left(1 + i\int_{t_0}^{t}dt'H_{el-c}(t')\right)n_I(t) \left(1 - i\int_{t_0}^{t}dt'H_{el-c}(t')\right)\nonumber\\
&\approx n_I(t) + i\int_{t_0}^{t}dt'\Big[H_{el-c}(t'), n_I(t)\Big]= n_I(t) + i\alpha\int_{t_0}^{t}dt'\Big[(a + a^\dag)n_I(t'), n_I(t)\Big]\,,
\label{noperator}
\end{align}
\end{widetext}
up to leading order in the coupling constant $\alpha$. Thus, the time-evolution of the electronic particle number contains, besides the electronic component, a contribution that arises because of the coupling to the cavity.  Introducing Eq.~\eqref{noperator} into Eq.~\eqref{eq_motion} we obtain:
\begin{align}
\left[H_{sys}, a\right]&= -\omega_c a -\alpha n = -\omega_c a -\alpha n_I(t)\nonumber\\
&- i\alpha^2\int_{t_0}^{t}dt'\Big[\left(a(t') + a^\dag(t')\right)n_I(t'), n_I(t)\Big].
\label{hsystem2}
\end{align}

Let us assume that $t_0 \rightarrow -\infty$ and write  Eq.~\eqref{hsystem2} as:
\begin{widetext}
\begin{align}
\left[H_{sys}, a\right]&= -\omega_c a -\alpha n_I(t)- i\alpha^2\int_{-\infty}^{t}dt'\Big[\Big(a(t)e^{-i\omega_c(t' - t)}+ a^\dag(t)e^{i\omega_c(t' - t)}\Big)n_I(t'), n_I(t)\Big]\,,
\label{hsystem3}
\end{align}
\end{widetext}
where $a(t)\approx ae^{-i\omega_ct}$  and $a^\dagger(t)\approx a^\dagger e^{i\omega_ct}$ in zeroth order in $\alpha$ (because the expression is already multiplied by $\alpha^2$ we can utilize the bare time dependence in this expression). In the following, we switch to the Fourier space in order to solve the equation for $a(t)$, and take into account all contributions that affect its  time-dependence, namely the external modes too. We obtain:
\begin{widetext}
\begin{align}
-i\omega a(\omega)&= -i\omega_c a(\omega) - \frac{\displaystyle \kappa}{\displaystyle 2}a(\omega) - \sqrt{\kappa}b_{in}(\omega) 
-i\frac{\displaystyle \alpha}{\displaystyle \hbar} n_I(\omega)\nonumber\\
&+ \frac{\displaystyle \alpha^2}{\displaystyle \hbar}\int_{-\infty}^{\infty}dt e^{i\omega t}\int_{-\infty}^{t}dt'\Big[\Big(a(t)e^{-i\omega_c(t' - t)}+ a^\dag(t)e^{i\omega_c(t' - t)}\Big)
n_I(t'), n_I(t)\Big]\,.
\label{aomega}
\end{align}
\end{widetext}
Before continuing with the derivation, let us describe each term in the above expression. The first term describes the free cavity evolution, the second term the leaking into the continuum of modes (the external bath) at rate $\kappa/2$, the third term is the input field supplied from the right side, the fourth term correspond to another ``input" contribution to the cavity from the electronic system (a noise term), while the last term leads to both a shift in the cavity frequency as well as to an extra decay channel ($Q$-factor modification).  One can now average  over the electronic system, thus neglecting any fluctuation (i.e. feed-back effects).  Moreover, we can neglect  the highly oscillating term $a^\dagger(t)\propto e^{i\omega_ct}$, namely we perform the so called Rotating Wave Approximation (RWA). Under all these assumptions, the last term in Eq.~\eqref{aomega} becomes:
\begin{align}
&\frac{\displaystyle \alpha^2}{\displaystyle \hbar}\int_{-\infty}^{\infty}dt e^{i\omega t}\int_{-\infty}^{t}dt'a(t)e^{-i\omega_c(t' - t)}\langle[n_I(t'), n_I(t)]\rangle_0\nonumber\\
&= i\frac{\displaystyle \alpha^2}{\displaystyle \hbar}\int_{-\infty}^{\infty}dt e^{i\omega t}a(t)\int_{-\infty}^{\infty}dt'(-i)\theta(t' - t)e^{-i\omega_c(t' - t)}\nonumber\\
&\times\langle[n_I(t'), n_I(t)]\rangle_0
= i\int_{-\infty}^{\infty}dt e^{i\omega t}a(t)\Pi(-\omega_c)\nonumber\\
&= ia(\omega)\Pi(-\omega_c),
\label{aomegalastterm2}
\end{align}
where
\begin{align}
\Pi(t' - t) = -i\theta(t' - t)\frac{\displaystyle \alpha^2}{\displaystyle \hbar}\langle[n_I(t'), n_I(t)]\rangle_0\,,
\label{corrfunc}
\end{align}
is the retarded density-density electronic correlation function utilized in the main text, and $\langle\dots\rangle_0$ means the expectation value of the unperturbed electronic system. Note that for deriving the above expression we assumed that the electronic system is in equilibrium and thus the time dependence is homogeneous. 

We are now in position to find the cavity field $a(t)$ and the output field $b_{out}(t)$ in terms of the input field $b_{in}(t)$. Introducing Eq.~\eqref{aomegalastterm2} into Eq.~\eqref{aomega} we obtain
\begin{align}
-i\omega a(\omega)&= -i\omega_c a(\omega) - \frac{\displaystyle \kappa}{\displaystyle 2}a(\omega) - \sqrt{\kappa}b_{in}(\omega) -i\frac{\displaystyle \alpha}{\displaystyle \hbar}\langle n_I(\omega)\rangle_0\nonumber\\
&+ ia(\omega)\Pi(-\omega_c)\,,
\label{aomega2}
\end{align}
so that 
\begin{align}
a(\omega) = -\frac{\displaystyle \sqrt{\kappa}b_{in}(\omega)+i(\alpha/\hbar)\langle n_I(\omega)\rangle_0}{\displaystyle -i(\omega - \omega_c) + \kappa/2 - i\Pi(-\omega_c)}\,.
\label{Res1}
\end{align}
We thus have two contributions to the cavity field: the external input and the input from the electronic system. Our aim is to relate in fact the output and input fields, which can be done easily via the expression in Eq.~\eqref{input_output}:
\begin{align}
&b_{out}(\omega)=\nonumber\\
&\frac{\displaystyle  [-i(\omega - \omega_c) - \kappa/2 - i\Pi(-\omega_c)]b_{in}(\omega)-i(\sqrt{\kappa}\alpha/\hbar)\langle n_I(\omega)\rangle_0}{\displaystyle -i(\omega - \omega_c) + \kappa/2 - i\Pi(-\omega_c)}\,.
\end{align}
In the limit of large number of photons in the input beam, we can neglect the contribution from the electronic system so that we obtain:
\begin{align}
b_{out}(\omega)&\approx \frac{\displaystyle -i(\omega - \omega_c) - \kappa/2 - i\Pi(-\omega_c)}{\displaystyle -i(\omega - \omega_c) + \kappa/2 - i\Pi(-\omega_c)}b_{in}(\omega)\,.
\end{align}

In experiments, one actually encounters a two-sided cavity (see Fig.~\ref{sketch} for the nomenclature), in which case the expression for the cavity equation of motion reads:
\begin{align}
\dot{a} = \frac{\displaystyle i}{\displaystyle \hbar}\left[H_{sys}, a\right] - \left(\frac{\displaystyle \kappa_1}{\displaystyle 2}+\frac{\displaystyle \kappa_2}{\displaystyle 2}\right)a - \sqrt{\kappa_1}b_{in}-\sqrt{\kappa_2}c_{in}\,,
\label{ain2}
\end{align}
so that for the output fields we get:
\begin{align}
b_{out}=\sqrt{\kappa_1}a+b_{in}\\
c_{out}=\sqrt{\kappa_2}a+c_{in}\,.
\end{align}
By following the same reasoning as for the one-sided cavity, we obtain: 
\begin{align}
a(\omega) = -\frac{\displaystyle \sqrt{\kappa_1}b_{in}(\omega) + \sqrt{\kappa_2}c_{in}(\omega)+i(\alpha/\hbar)\langle n_I(\omega)\rangle_0}{\displaystyle -i(\omega - \omega_c) + \kappa_1/2 + \kappa_2/2- i\Pi(-\omega_c)},
\label{res1twosided}
\end{align}
while if the two mirrors are the same $\kappa_1 = \kappa_2\equiv\kappa$, this becomes
\begin{align}
a(\omega) = -\frac{\displaystyle \sqrt{\kappa}[b_{in}(\omega) + c_{in}(\omega)]+i(\alpha/\hbar)\langle n_I(\omega)\rangle_0}{\displaystyle -i(\omega - \omega_c) + \kappa- i\Pi(-\omega_c)}\,.
\label{res2twosided}
\end{align}
Assuming again that the input flux is much larger than the electronic contribution, we can write:
\begin{equation}
c_{out}(\omega)=-\tau b_{in}(\omega) + (\ldots) c_{in} (\omega)
\end{equation}
with
\begin{equation}
{\bf \tau}=\frac{\kappa}{{\displaystyle -i(\omega - \omega_c) + \kappa- i\Pi(-\omega_c)}}\equiv Ae^{i\phi}
\end{equation}
being the transmission of the cavity, which is a complex number, and which depends on the electronic susceptibility $\Pi(\omega)$, as stated in Eq.(\ref{transmissioncoef}) in the MT. 

\section{The susceptibility for the Kitaev model in case of periodic boundary conditions}\label{bulksusceptibility}

In this section we give more details on the derivation of the susceptibility $\Pi(\omega)$ for the case of a ring geometry for which we can apply PBCs. The Kitaev chain Hamiltonian in real space was defined in Eq.(\ref{kitaevhamiltonian}) as (for definitions of the parameters please see MT):
\begin{align} 
H_{tot}&=H_{el}+H_{el-c}+H_{ph}\,,\\
H_{el} &=\!- \mu\sum\limits_{i = 1}^{N}c^\dag_ic_i-\frac{1}{2}\sum\limits_{i = 1}^{N - 1}({\rm t}c^\dag_ic_{i + 1} \!+\Delta c_ic_{i + 1} \!+ {\rm h.c.})\nonumber\,,\\
H_{int}&= \alpha\sum\limits_{i = 1}^{N}c^\dag_ic_i(a + a^\dag)\,,
\end{align}
and $H_{ph}=\omega_ca^\dagger a$. In this case, we can switch to the Fourier space, which implies we can write the fermionic operators as follows:  
\begin{equation}
c_j = \frac{\displaystyle 1}{\displaystyle \sqrt{N}}\sum_k c_k e^{ikj}\,.
\end{equation}
where $c_k$ is the fermionic annihilation operator with momentum $k=2\pi n/N$. We can then rewrite the electronic Hamiltonian in momentum space 
\begin{equation}
H_{el} = \sum_{k > 0}H^K_{BdG}(k)\,,
\end{equation}
with
\begin{equation}
H^k_{BdG}(k) = \xi_k(c^\dag_k c_k - c_{-k}c^\dag_{-k}) -i\Delta\sin{k}(c_{-k} c_k - c^\dag_{k}c^\dag_{-k}),
\label{1dhamiltoniankspace}
\end{equation}
is the Bogoliubov de Gennes Hamiltonian and  $\xi_k = -{\rm t}\cos{k} - \mu$.
The interaction Hamiltonian between the electronic system and the cavity can as well be written in the $k$-space as
\begin{equation}
H_{el-c} = \sum_{k > 0} \alpha (c^\dag_k c_k - c_{-k}c^\dag_{-k})(a+a^\dagger).
\label{interactionhamiltoniankspace}
\end{equation}
One can simply diagonalize the $H_{el}$ in the $k$-space and write:
\begin{equation}
H_{el} = \sum_{k>0}E_k\left(\gamma_k^\dag\gamma_k + \gamma_{-k}^\dag\gamma_{-k}\right),
\label{quadratichamiltonianb}
\end{equation}
with
\begin{equation}
E_{k} = \pm\sqrt{(-{\rm t}\cos{k} - \mu)^2 + (\Delta \sin k)^2}\,,
\label{energyspectrum}
\end{equation}
being the eigenenergies, and we used 
\begin{equation}
c_k = u_k^*\gamma_k + v_k\gamma_{-k}^\dag,
\label{bogoliubovtransform1}
\end{equation}
\begin{equation}
c_{-k}^\dag = -v_k^*\gamma_k + u_k\gamma_{-k}^\dag\,.
\label{bogoliubovtransform2}
\end{equation}
Here,  the functions $u_k$ and $v_k$ are given by
\begin{equation}
|u_k|^2 = \frac{\displaystyle 1}{\displaystyle 2}\left(1 + \frac{\displaystyle \xi_k}{\displaystyle E_k}\right)\,,
\label{uk}
\end{equation}
\begin{equation}
|v_k|^2 = \frac{\displaystyle 1}{\displaystyle 2}\left(1 - \frac{\displaystyle \xi_k}{\displaystyle E_k}\right).
\label{vk}
\end{equation}
We can now calculate the susceptibility $\Pi(\omega)$ defined in the previous section, which quantifies the change in the photonic transmission $\tau$ due to the interaction with the electronic system. In the Fourier space, we obtain:
\begin{align}
\Pi(t)&= -i\theta(t)\alpha^2\sum_{k, q>0}\langle\Big[\left(c^\dag_k c_k - c_{-k}c^\dag_{-k}\right)(t),\nonumber\\
&\left(c^\dag_q c_q - c_{-q}c^\dag_{-q}\right)(0)\Big]\rangle.
\label{correlationfunckspace}
\end{align}
Using Eq.~\eqref{bogoliubovtransform1} and Eq.~\eqref{bogoliubovtransform2} with Eq.~\eqref{uk} and Eq.~\eqref{vk} we find:
\begin{equation}
\Pi(\omega) = \alpha^2\sum_{k > 0} \frac{\displaystyle \left(\Delta\sin k\right)^2}{\displaystyle E_{k}^2}\left(\frac{\displaystyle 1}{\displaystyle \omega - 2E_{k} + i\eta} - \frac{\displaystyle 1}{\displaystyle \omega + 2E_{k} + i\eta}\right),
\end{equation}
with the small  $\eta>0$ assuring the convergence of the time-integrals. For large $N\gg1$, we can transform the sum into integral, and also write $\Pi(\omega)=\Pi'(\omega)+i\Pi''(\omega)$, with:
\begin{align}
\Pi'(\omega)&=\frac{2N\alpha^2}{\pi}\mathcal{P}\int_0^{\pi}dk\frac{\displaystyle \left(\Delta\sin k\right)^2}{\displaystyle E_{k}}\frac{\displaystyle 1}{\displaystyle \omega^2 - 4E_{k}^2}\,,\\
\Pi''(\omega)&=\frac{N\alpha^2}{2}\int_0^{\pi}dk\frac{\displaystyle \left(\Delta\sin k\right)^2}{\displaystyle E_{k}^2}\left[\delta(\omega-2E_k)-\delta(\omega+2E_k)\right]\,,
\end{align}
where $\mathcal{P}\dots$ means the principal value of the function and we used the fact that:
\begin{equation}
\frac{1}{x-a+i\epsilon}=\mathcal{P}\frac{1}{x-a}-i\pi\delta(x-a)\,.
\end{equation}
We can perform the integral over $k$ for the imaginary part $\Pi'(\omega)$ to obtain the expression in the MT:
\begin{align}
\Pi''(\omega) = \frac{\alpha^2 {\rm t} N}{2\mu\omega}\sqrt{1 - \frac{\left[\left(\omega/2\right)^2 - {\rm t}^2 - \mu^2\right]^2}{\displaystyle 4{\rm t}^2\mu^2}}\,,
\end{align}
for $\left|{\rm t} + \mu\right| < \omega/2 < \left|{\rm t} - \mu\right|$, and being zero otherwise. For the real part $\Pi'(\omega)$ we found no simple solution, and so we chose not to depict it. Note that the susceptibility $\Pi(\omega)\propto N$, i. e. it scales linearly with the number of sites.   

\section{The susceptibility of the nanowire in case of periodic boundary conditions}\label{nanowire_periodic}

The Hamiltonian for a nanowire in the presence of a Zeeman field $B$, Rashba spin-orbit coupling $u$ and induced superconductivity $\Delta$ reads \cite{pientka2013magneto} 

\begin{align}
H_{NW}=\left(\frac{\displaystyle p^2}{\displaystyle 2m}-\mu\right)\tau_z+up\sigma_z\tau_z+B\sigma_x+\Delta\tau_x.
\end{align}

Let us diagonalize the Hamiltonian $H_{NW}=H_0+\Delta\tau_x$ in the absence of the induced pairing $\Delta$ and then treat the latter perturbatively
\begin{align}
H_{0}=\left(\frac{\displaystyle p^2}{\displaystyle 2m}-\mu\right)\tau_z+up\sigma_z\tau_z+B\sigma_x.
\end{align}
In order to do so, let us perform the unitary transformation~\cite{pientka2013magneto}

\begin{align}
\tilde{H}_{0}=UH_{0}U^\dag,
\end{align}
where
\begin{align}
U=\exp\left(i\alpha\sigma_y\tau_z/2\right)=\cos\left(\alpha/2\right)+i\sigma_y\tau_z\sin\left(\alpha/2\right).
\end{align}

If we choose $\alpha$ so that
\begin{align}
\tan\left(\alpha\right)=\frac{\displaystyle B}{\displaystyle up},
\end{align}
 $\tilde{H}_{0}$ takes the from




\begin{align}
\tilde{H}_{0}=\Big[\frac{\displaystyle p^2}{\displaystyle 2m}-\mu+\sqrt{u^2p^2+B^2}\sigma_z\Big]\tau_z.
\end{align}

Now let us reintroduce the pairing term $\Delta\tau_x$ and apply the transformation $U$ to it.


Then $\tilde{H}_{NW}$ reads
\begin{align}
\tilde{H}_{NW}&=\Big[\frac{\displaystyle p^2}{\displaystyle 2m}-\mu+\sqrt{u^2p^2+B^2}\sigma_z\Big]\tau_z+\frac{\displaystyle \Delta up}{\displaystyle \sqrt{u^2p^2+B^2}}\tau_x\nonumber\\
&-\frac{\displaystyle \Delta B}{\displaystyle \sqrt{u^2p^2+B^2}}\sigma_y\tau_y.
\label{htransnw}
\end{align}

Neglecting the last term in Eq.(\ref{htransnw}) the effective Hamiltonian reads
\begin{align}
\tilde{H}_{\rm eff}=\Big[\frac{\displaystyle p^2}{\displaystyle 2m}-\mu+\sqrt{u^2p^2+B^2}\sigma_z\Big]\tau_z+\frac{\displaystyle \Delta up}{\displaystyle \sqrt{u^2p^2+B^2}}\tau_x.
\label{heff}
\end{align}


The bulk energy spectrum of $\tilde{H}_{\rm eff}$ for $\sigma_z=-1$ is

\begin{align}
\epsilon_{p,-1}=\pm\sqrt{\left(\frac{\displaystyle p^2}{\displaystyle 2m}-\mu-\sqrt{u^2p^2+B^2}\right)^2+\frac{\displaystyle \Delta^2 u^2p^2}{\displaystyle u^2p^2+B^2}}
\end{align}

and for $\sigma_z=+1$ is

\begin{align}
\epsilon_{p,+1}=\pm\sqrt{\left(\frac{\displaystyle p^2}{\displaystyle 2m}-\mu+\sqrt{u^2p^2+B^2}\right)^2+\frac{\displaystyle \Delta^2 u^2p^2}{\displaystyle u^2p^2+B^2}}.
\end{align}

The electronic susceptibility is defined as
\begin{align}
\Pi(t)=-i\theta(t)\langle\left[\tau_z(t), \tau_z(0)\right]\rangle.
\end{align}

In order to diagonalize the Hamiltonian in Eq.~(\ref{heff}), let us perform a transformation 
\begin{align}
H_d=U_1\tilde{H}_{eff}U_1^\dag=\epsilon_p\tau_z,
\end{align}
where
\begin{align}
U_1=\exp\left(i\gamma\tau_y/2\right)=\cos\left(\gamma/2\right)+i\tau_y\sin\left(\gamma/2\right).
\end{align}





Then the electronic susceptibility reads
\begin{align}
\Pi(t)&=-i\theta(t)\langle\Big[\cos(\gamma)\tau_z(0)-\sin(\gamma)\tau_x(t),\nonumber\\
&\cos(\gamma)\tau_z(0)-\sin(\gamma)\tau_x(0)\Big]\rangle.
\label{pit}
\end{align}

\begin{align}
\tau_x(t)&=e^{i\epsilon_p\tau_zt}\tau_x(0)e^{-i\epsilon_p\tau_zt}\nonumber\\
&=
\cos\left(2\epsilon_pt\right)\tau_x(0)-\sin\left(2\epsilon_pt\right)\tau_y(0).
\label{tauxtime}
\end{align}

Introducing Eq.(\ref{tauxtime}) into Eq.(\ref{pit}), $\Pi(t)$ reads

\begin{align}
\Pi(t)
=-2\theta(t)\sin^2(\gamma)\sin(2\epsilon_p t).
\label{pit2}
\end{align}

Performing the Fourier transform, Eq.~(\ref{pit2}) reads

\begin{align}
\Pi(\omega)&=-2\sum_p\sin^2(\gamma)\int_{0}^{+\infty}dt e^{i\omega t-\eta t}\sin(2\epsilon_p t)\nonumber\\
&=-2\sum_p\sin^2(\gamma)\frac{\displaystyle 2\epsilon_p}{\displaystyle 4\epsilon_p^2+(\eta-i\omega)^2}\nonumber\\
&=\sum_p\sin^2(\gamma)\left(\frac{\displaystyle 1}{\displaystyle \omega-2\epsilon_p+i\eta}-\frac{\displaystyle 1}{\displaystyle \omega+2\epsilon_p+i\eta}\right).
\label{suscept}
\end{align}





Sum can be transformed into integral as
\begin{align}
\sum_p=\frac{\displaystyle N}{\displaystyle 2\pi}\int dp=\frac{\displaystyle N}{\displaystyle 2\pi}\int d\epsilon \rho(\epsilon).
\end{align}

Then the imaginary part of the susceptibility $\Pi(\omega)=\Pi'(\omega)+i\Pi''(\omega)$, Eq.~(\ref{suscept}) reads
\begin{align}
\Pi''(\omega)=-\frac{\displaystyle N}{\displaystyle 4}\int d\epsilon \rho(\epsilon)\sin^2(\gamma)(\epsilon)\left[\delta(\epsilon-\omega/2)-\delta(\epsilon+\omega/2)\right].
\label{susceptgeneral}
\end{align}

\subsection{The projection onto the lower band $\sigma_z=-1$}
The low-energy subspace at $p=\pm p_F$ is formed by the bands for which $\sigma_z=-1$. $p_F=2mu$ when $\mu=0$.

The effective Hamiltonian for $\epsilon_{SO}=mu^2>>B$, $\mu=0$ linearized around $\pm p_F$ reads

\begin{align}
\tilde{H}_{eff}=u(|p|-p_F)\tau_z+sign(p)\Delta\tau_x.
\label{heffm1}
\end{align}

The bulk energy spectrum reads
\begin{align}
\epsilon_p=\pm\sqrt{\Delta^2+u^2(|p|-p_F)^2}.
\end{align}


In this limit
\begin{align}
\sin^2(\gamma)(\epsilon)=\frac{\displaystyle \Delta^2}{\displaystyle \epsilon^2}
\label{sin1}
\end{align}
and 
 the density of states reads
\begin{align}
\rho(\epsilon)=\frac{\displaystyle \epsilon }{\displaystyle u\sqrt{\epsilon^2-\Delta^2} sign(p)}.
\label{density1}
\end{align}

Introducing Eq.(\ref{sin1}) and Eq.(\ref{density1}) into Eq.(\ref{susceptgeneral}) the imaginary part of the electronic susceptibility reads
\begin{align}
\Pi''(\omega)=-\frac{\displaystyle N\Delta^2}{\displaystyle \omega u\sqrt{\omega^2-4\Delta^2}}.
\end{align}









\subsection{The projection onto $\sigma_z=+1$}
In this case the effective Hamiltonian linearized around $p=0$ reads

\begin{align}
\tilde{H}_{eff}
=\left(\frac{\displaystyle p^2}{\displaystyle 2m^*}-\mu+B\right)\tau_z+\frac{\displaystyle \Delta u p}{\displaystyle B}\tau_x,
\label{heffp2}
\end{align}

where an effective mass $m^*$ was introduced as
\begin{align}
\frac{\displaystyle 1}{\displaystyle m^*}=\frac{\displaystyle 1}{\displaystyle m}+\frac{\displaystyle u^2}{\displaystyle B}.
\end{align}



The bulk energy spectrum reads
\begin{align}
\epsilon_p&=\pm\sqrt{\left(B-\mu+\frac{\displaystyle p^2}{\displaystyle 2m^*}\right)^2+\frac{\displaystyle \Delta^2 u^2 p^2}{\displaystyle B^2}}\nonumber\\
&\approx \pm\sqrt{ (B-\mu)^2 + cp^2},
\label{energyapp2}
\end{align}
where

\begin{align}
c=\frac{\displaystyle B-\mu}{\displaystyle m^*}+\frac{\displaystyle \Delta^2 u^2}{\displaystyle B^2}.
\end{align}



The density of state reads
\begin{align}
\rho(\epsilon)=\frac{\displaystyle \epsilon}{\displaystyle \sqrt{c\left[\epsilon^2-(B-\mu)^2\right]}}.
\label{density2}
\end{align}
And
\begin{align}
\sin(\gamma)^2(\epsilon)=\frac{\displaystyle \Delta^2 u^2\left[\epsilon^2-(B-\mu)^2\right]}{\displaystyle B^2\epsilon^2c}.
\label{sin2}
\end{align}

Introducing Eq.(\ref{density2}) and Eq.(\ref{sin2}) into Eq.(\ref{susceptgeneral}) the imaginary part of the electronic susceptibility reads
\begin{align}
\Pi''(\omega)=-\frac{\displaystyle N\Delta^2 u^2 \sqrt{\omega^2-4(B-\mu)^2}}{4\omega B^2 c^{3/2}}.
\label{susceptapp2}
\end{align}

\section{The correlation function for open boundary conditions}\label{openboundaries}

In this part, we present details on the calculation of the susceptibility for a finite wire with open boundary conditions. In this case, the excitation spectrum changes compared to the previous case, as in the topological region the zero energy Majorana fermions emerge. In the MT we found that the susceptibility reads:
\begin{equation}
\Pi(\omega)=-i\alpha^2\sum_{i,j=1}^N\int_0^\infty dte^{-i\omega t}\langle|[\hat{n}_i(t),\hat{n}_j(0)]|\rangle
\label{suscept2}
\end{equation}
and which can  be written as
\begin{equation}
\Pi(\omega)=\Pi_{BB}(\omega)+\Pi_{BM}(\omega)\,,
\end{equation}
being the sum of a bulk susceptibility, that can be constructed from only the bulk (or gaped) states, and cross terms that involve both bulk and Majorana states, respectively.  We use the discrete lattice model to numerically diagonalize the Hamiltonian for an electronic system with $N$ fermionic sites. 
This can be written in a compact form as follows:  
\begin{align}
H_{el} = \frac{\displaystyle 1}{\displaystyle 2}\vec{c}^\dag M \vec{c}
\end{align}
with
\begin{align}
\vec{c} = (c_1, c_1^\dag, c_2, c_2^\dag, ..., c_N, c_N^\dag)^T\,,
\end{align}
where $M$ is a $2N\times2N$ matrix. Moreover, assuming all entries in the matrix  are real,  $M$ is also symmetric, we can write it as follows: 
\begin{align}
M = PW P^\dag,
\end{align}
where $W$ is a diagonal matrix with eigenvalues on its diagonal and $P$ is a unitary matrix ($P P^\dag = P^\dag P = I$) whose columns are eigenvectors of $M$. The matrix $W$ is ordered so that 
\begin{align}
W = 
\begin{pmatrix}
\epsilon_1 & 0 & \cdots & 0 & 0 \\
0 & -\epsilon_1 & \cdots & 0 & 0\\
\vdots & \vdots & \ddots & \vdots & \vdots\\
0 & 0 & \cdots & \epsilon_N & 0\\
0 & 0 & \cdots & 0 & -\epsilon_N\,
\end{pmatrix}
\,,
\end{align}
with  $\pm\epsilon_{n}$ being the eigenenergies of the BdG Hamiltonian and $n=1,\dots N$. That pertains to the following diagonal Hamiltonian:  
\begin{align}
H_{1D} = \sum_{m = 1}^{N}\epsilon_m\left(\tilde{c}_m^\dag\tilde{c}_m - \frac{\displaystyle 1}{\displaystyle 2}\right)\,,
\label{diagonalh}
\end{align}
where $\tilde{c}_m$ ($\tilde{c}_m^\dagger$) are the annihilation (creation) operators for the Bogoliubov quasiparticles, which are defined as follows:
\begin{align}
\vec{\tilde{c}}=P^\dagger\vec{c}
\end{align}
and 
\begin{align}
\vec{\tilde{c}}= (\tilde{c}_1,\tilde{c}_1^\dag, \tilde{c}_2, \tilde{c}_2^\dag,\dots, \tilde{c}_N, \tilde{c}_N^\dag)^T\,.
\end{align}
It is instructive to introduce the wavefunctions $\vec{\psi}_k\left(i\right) = \left(u_{k, i}, v_{k, i}\right)^T$, where $u_{k, i}\left(v_{k, i}\right) = p_{2i - 1, k}\left(p_{2i, k}\right)$, and which are describing the state $k=0,\dots,2N$ at position $i=1,\dots,N$ in the lattice and accounts for the electron ($u$) and hole ($v$) components, respectively. That allows us to write:
\begin{align}
c_i = \sum_{k = 1}^{N}\left[u_{2k-1,i}\tilde{c}_k + u_{2k, i}\tilde{c}_k^\dag\right],
\label{ci}
\end{align}
\begin{align}
c_i^\dag = \sum_{k = 1}^{N}\left[v_{2k-1,i}\tilde{c}_k + v_{2k,i}\tilde{c}_k^\dag\right].
\label{cidag}
\end{align}
so that we can rewrite Eq.~\eqref{suscept2} in terms of $u_{k, i}$ and $v_{k, i}$ as follows:
\begin{align}
&\Pi(\omega)= \sum_{i,j,k,m = 1}^{N} (1 - n_k)(1 - n_m)v_{2k - 1, i}u_{2m - 1, i}\nonumber\\
&\times\left(u_{2k, j}v_{2m, j} - v_{2k, j}u_{2m, j}\right)\nonumber\\
&\times\left(\frac{1}{\omega + i\eta - \epsilon_{2k - 1} - \epsilon_{2m - 1}}-\frac{1}{\omega + i\eta + \epsilon_{2k - 1} + \epsilon_{2m - 1}}\right)\nonumber\\
&+ (1 - n_k)n_m v_{2k - 1, i}u_{2m, i}\left(u_{2k, j}v_{2m - 1, j} - v_{2k, j}u_{2m - 1, j}\right)\nonumber\\
&\times\left(\frac{1}{\omega + i\eta - \epsilon_{2k - 1} + \epsilon_{2m - 1}}-\frac{1}{\omega + i\eta + \epsilon_{2k - 1} - \epsilon_{2m - 1}}\right) \nonumber\\
&+n_k(1 - n_m) v_{2k, i}u_{2m - 1, i}\left(u_{2k - 1, j}v_{2m, j} - v_{2k - 1, j}u_{2m, j}\right)\nonumber\\
&\times\left(\frac{1}{\omega + i\eta + \epsilon_{2k - 1} - \epsilon_{2m - 1}} - \frac{1}{\omega + i\eta - \epsilon_{2k - 1} + \epsilon_{2m - 1}}\right) \nonumber\\
&+ n_kn_m v_{2k, i}u_{2m, i}\left(u_{2k - 1, j}v_{2m - 1, j} - v_{2k - 1, j}u_{2m - 1, j}\right)\nonumber\\
&\times\left(\frac{1}{\omega + i\eta + \epsilon_{2k - 1} + \epsilon_{2m - 1}} 
- \frac{1}{\omega + i\eta - \epsilon_{2k - 1} - \epsilon_{2m - 1}}\right).
\end{align}
Next we extract from this expression only the $\Pi_{BM}(\omega)$ component. This reads:
\begin{align}
&\Pi_{BM}(\omega)= \sum_{i,j = 1}^{N}\sum_{k = 1}^{N - 1}\Bigg(\frac{1}{\omega + i\eta + \epsilon_{2k - 1}} + \frac{1}{-\omega - i\eta + \epsilon_{2k - 1}}\Bigg)\nonumber\\
&\times\Bigg[ -(1 - n_k - n_M)\Big[v_{2k - 1, i}u_{2M - 1, i}u_{2k, j}v_{2M, j}\nonumber\\ 
&- v_{2k - 1, i}u_{2M - 1, i}v_{2k, j}u_{2M, j}
+ v_{2M - 1, i}u_{2k - 1, i}u_{2M, j}v_{2k, j}\nonumber\\
&- v_{2M - 1, i}u_{2k - 1, i}v_{2M, j}u_{2k, j}\Big] \nonumber\\
&-(n_M - n_k)\Big[v_{2k - 1, i}u_{2M, i}u_{2k, j}v_{2M - 1, j}\nonumber\\
&- v_{2k - 1, i}u_{2M, i}v_{2k, j}u_{2M - 1, j}+ v_{2M, i}u_{2k - 1, i}u_{2M - 1, j}v_{2k, j}\nonumber\\
&- v_{2M, i}u_{2k - 1, i}v_{2M - 1, j}u_{2k, j}\Big]\Bigg]\,,
\label{suscept4}
\end{align}
and, using that $u_{2k - 1, i} = v_{2k, i}$, it can be simplified even further to give:
\begin{align}
&\Pi_{BM}(\omega)= \sum_{i,j = 1}^{N}\sum_{k = 1}^{N - 1}\Bigg(\frac{1}{\omega + i\eta + \epsilon_{2k - 1}} + \frac{1}{-\omega - i\eta + \epsilon_{2k - 1}}\Bigg)\nonumber\\
&\times\Bigg[ -(n_M - 1 + n_k)\Big[-u_{2k, i}v_{2M, i}u_{2k, j}v_{2M, j}\nonumber\\
&+ u_{2k, i}v_{2M, i}v_{2k, j}u_{2M, j}
- u_{2M, i}v_{2k, i}u_{2M, j}v_{2k, j}\nonumber\\
 &+ u_{2M, i}v_{2k, i}v_{2M, j}u_{2k, j}\Big] \nonumber\\
&+(n_M - n_k)\Big[-u_{2k, i}u_{2M, i}u_{2k, j}u_{2M, j}\nonumber\\
&+ u_{2k, i}u_{2M, i}v_{2k, j}v_{2M, j} 
- v_{2M, i}v_{2k, i}v_{2M, j}v_{2k, j}\nonumber\\
&+ v_{2M, i}v_{2k, i}u_{2M, j}u_{2k, j}\Big]\Bigg].
\end{align}
Let us now introduce the coefficients $C^{(s)}$, $s = 1,2$ defined in the MT:
\begin{align}
C^{(s)}&= \sum_{i = 1}^{N}(u_{2M, i}\delta_{s, 1} + v_{2M, i}\delta_{s, 2})u_{2k, i}\nonumber\\ 
&- (u_{2M, i}\delta_{s, 2} + v_{2M, i}\delta_{s, 1})v_{2k, i}\,,
\end{align}
which we can utilize to rewrite $\Pi_{BM}(\omega)$ as follows:
\begin{align}
&\Pi_{BM}(\omega)= \sum_{k = 1}^{N - 1}\Bigg(\frac{1}{\epsilon_{2k} + \omega + i\eta} + \frac{1}{\epsilon_{2k} - \omega - i\eta}\Bigg)\nonumber\\
&\times\Bigg[(n_M - n_k)\left|C^{(1)}\right|^2 - (n_M - 1 + n_k)\left|C^{(2)}\right|^2\Bigg]\,,
\end{align}
and which correspond to the expression Eq.~(\ref{pibm}) in the MT.

%

\end{document}